\documentclass[structabstract]{aa}
\usepackage{graphicx}
\usepackage{txfonts}
\usepackage{amssymb}
\usepackage{epstopdf}
\usepackage{fancyhdr}
\usepackage{./natbib}
\bibpunct{(}{)}{;}{a}{}{,} 

\def \XMM {{\it XMM-Newton}}
\def  \swi {{\it Swift}}
\def \rchisq {$\chi_{\nu} ^{2}$}
\def \chisq {$\chi^{2}$}
\begin{document}
   \title{A Reanalysis of High Resolution \emph{XMM-Newton} Data of V2491 Cyg Using Collisionally Ionized Hot Absorber Models}

     \author{\c{S}. Balman 
          \inst{1} \thanks{solen@astroa.physics.metu.edu.tr}
          \and
	  \c{C}. Gams{\i}zkan
	  \inst{1}  \thanks{cigdem.gamsizkan@gmail.com}
          }
\institute{Department of Physics, Middle East Technical University, Dumlup{\i}nar Bulvar{\i} 06800 Ankara, Turkey
             }

   \date{Received -; Accepted -}
    \authorrunning{Balman \& Gams{\i}zkan}
   \titlerunning{Reanalysis of V2491 Cyg RGS Data with Hot Absorber Models}

  \abstract
   {}
   { Modeling of absorption features in the high resolution \XMM\ Reflection
Grating Spectrometer data.
}
   { X-ray spectral analysis and modeling using the SRON (HEA Division, Netherlands Institute for
Space Research) software SPEX  version 2.05.04\ .
}
   { We present a reanalysis of \XMM\ Reflection Grating Spectrometer
data of the classical nova V2491 Cyg obtained from two different pointings, 40 d and 50 d after outburst.  
We aim to model absorption components
in the high resolution spectra independently from the continuum
model. In order to model the complex  absorption, we are 
utilizing hot collisionally ionized absorber models along with interstellar absorption (of gas and dust origin separately)
that we discuss in the light of observations.
For an adequate approximation, and ease of fitting procedures we use a blackbody model for the 
continuum. 
We find blackbody temperatures in a range 61-91 eV yielding a white dwarf (WD)  mass
of 1.15-1.3 M$_{\odot}$ assuming this range is the maximum temperature achieved during the H-burning phase.   
We derive two different hot absorber components from our fits  with blueshifts yielding 2900-3800 km s$^{-1}$ for the first
(day 40) 
and  2600-3600 km s$^{-1}$ for the second observation 50 days after outburst consistent with ejecta/wind speeds.
The two collisionally ionized hot absorption components have temperatures
kT$_{\rm 1}\simeq1.0-3.6$ keV and kT$_{\rm 2}\simeq0.4-0.87$ keV with $rms$ velocities $\sigma_{\rm v1}\sim872$ km/s and  $\sigma_{\rm v2}\sim56$ km/s.
These are consistent with shock temperatures in the X-ray wavelengths.
V2491 Cyg shows signature of H-burning with
underabundant carbon C/C$_{\odot}$=0.3-0.5, and enhanced nitrogen N/N$_{\odot}$=5-7  and  oxygen O/O$_{\odot}$=16-43. The high oxygen overabundance hints at a C-O WD.
We find the equivalent hydrogen column density of the hot collisionally ionized (in equilibrium) absorbers
in a range  (0.6-18.0)$\times$10$^{23}$ cm$^{-2}$ and  (2.0-5.3)$\times$10$^{23}$ cm$^{-2}$ on days 40 and 
50 after outburst, respectively.
Our fits yield the most adequate \rchisq\ (range 1.8-2.9) up-to-date obtained for the modeling 
of high resolution X-ray data of V2491 Cyg. An additional photoionized absorber (third intrinsic absorber component) originating
in the shell/ejecta improves the model fits with  \rchisq\ in a range 1.7-2.5, but shows only (1-0.1)\%\ of the absorption
by the collisionaly-ionized hot gas. Our analysis reveals a second blackbody component on day 50 with effective temperature 120-131 eV and
effective radius about 10\%\ of the WD which may indicate the onset of magnetic accretion.}
{}

\maketitle


\section{Introduction}

Classical novae (CNe) outbursts occur in cataclysmic variable (CV) systems
on the surface of the WD as a result of an explosive ignition of
accreted material (Thermonuclear Runaway-TNR) causing
the ejection of 10$^{-7}$ to 10$^{-3}$ M$_{\odot}$ of material at velocities
up to several thousand kilometers per second \citep{1989PASPshara, 1994inbilivio, 2012BASIstarrfield, 2008bookbode}.

After the initial expansion phase of the outburst,
the velocity of the material in
deeper zones of the envelope drops quickly and a hydrostatic equilibrium is
established \citep{1989PASPshara, 2008bookbode}. A gradual hardening of the
stellar remnant spectrum with time past visual maximum is expected  consistent with H-burning at constant bolometric luminosity and
decreasing photospheric radius, as the envelope mass is depleted. This residual
hydrogen-rich envelope matter is mainly consumed by H-burning and wind-driven
mass loss. The emission from the remnant WD is a blackbody-like stellar continuum.
As the stellar photospheric radius decreases in time during the constant bolometric
luminosity phase, the
effective photospheric temperature increases
(up to values in the range 1--10 $\times 10^5$ K) and the peak of
the stellar spectrum is shifted from visual to ultraviolet and
to the X-ray energy band, where finally the H-burning turns off
\citep[e.g.,][]{1991ApJmacdonald, 1998ApJbalman, 1999AAkahabka, 2001MNRASbalman, 2003ApJness, 2002MNRASorio, 2003ApJorio, 2007ApJness,
2008ApJnelson, 2010MNRASpage, 2010ApJrauch, 2011ApJosborne, 2013ApJtofflemire, 2014AAhenze, 2015MNRASpage}.
This constitutes the soft X-ray component (0.1-1.0 keV, 124.0-12.4 \AA ) detected during the nova outbursts.

CNe are, also, detected in the hard X-rays (above 0.5 keV, shorter than 24.8 \AA ) {\it during the outburst stage} as a
result of shocked plasma emission in the accretion process, wind outflow,
wind-wind and/or blast wave interactions
\citep[e.g.,][]{1994MNRASobrien, 1996ApJkrautter, 1998ApJbalman, 2001ApJmukai, 2001MNRASorio, 2003ApJness, 2002Scihernanz, 2007ApJhernanz,
2006Natursokoloski, 2009AJness, 2010MNRASpage, 2014ApJchomiuk, 2015MNRASorio}.
Recurrent novae (RNe) are a type of CNe with outbursts
occurring at intervals of several decades \citep{1987ApJwebbink, 2001ApJhachisu, 2008bookbode}.

In this paper, we reanalyze the \XMM\ Reflection Grating Spectrometer (RGS) data of classical
nova V2491 Cyg. The high resolution nova spectra 
show the existence of absorption features in the X-ray wavelengths. 
We assume that there is complex absorption of interstellar, photospheric, and
of collisionally and/or photoionized gas origin due to moving material in   
the line of sight from a nova wind or ejecta.
Our main aim is to model the absorption components
detected in the high resolution spectra independently from the continuum
model and compare the photoionized warm and collisionally ionized hot absorber gas in this system. 
In addition to deriving the absorption properties, 
we obtain abundances of elements in the ejecta/nova wind.
In comparison with previous modeling of individual absorption features with
Gaussians, NLTE (Non-Local Thermodynamic Equilibrium) static atmosphere models, NLTE
expanding atmosphere models, or intrinsic photoionized warm 
absorber models, we are utilizing 
collisionally ionized absorber models that we discuss in the light of observations.
We will compare the results obtained  
with the same \XMM\ RGS data (we use here)
using photoionized warm absorber models as described in Pinto et al. (2012)
with the hot collisionally ionized (in equilibrium) absorber model fits for V2491 Cyg using SPEX software.
We caution that we use blackbody models for the continuum (as was used in Pinto et al. 2012), 
thus we will be missing some photospheric absorption features.

Nova 2491 Cyg was discovered in April 2008 at about 7.7 mag on
CCD frames in white light \citep{2008IAUCnakano}. \citet{2011NEWAmunari}\ calculates
time to decay by 2 magnitudes of 4.8 days making this nova a very fast nova similar to
V838 Her and V2487 Oph.
V2491 Cyg has been classified as a He/N nova based on
photometric and spectroscopic results \citep{2008IAUClynch, 2008CBEThelton}. 
The spectra have very broad lines with complex profiles and large expansion
velocities ($\sim$4000-6000 km s$^{-1}$)  \citep{2008IAUClynch, 2008ATeltomov}.
V2491 Cyg shows both the hard X-ray and the soft X-ray components during the outburst stage
\citep[analysis can be found in][]{2010MNRASpage, 2011ApJness}.
\citet{2009AAibarra}\ show that V2491 was a persistent X-ray source using
archival $ROSAT$, \XMM, and \swi\ data during its quiescent phase before the
optical outburst. \citet{2011PASJtakei}\ reveal early nonthermal emission from the nova during the
outburst stage and find that accretion is re-established as early as day 50 (also plausibly day 40) after outburst.
Recently, observations of the source more than two years after the outburst indicate that the quiescent source 
is a bright hard X-ray emitter including blackbody emission with 77 eV effective temperature indicating
the characteristics of a soft Intermediate Polar (magnetic) type  CV \citep{2015ApJzemko}.  

\section{The Data and Observations}

The \XMM\ Observatory \citep{2001AAjansen}\ has three 1500 cm$^2$
X-ray telescopes each with an European Photon Imaging Camera (EPIC)
at the focus; two of which have Metal Oxide Semiconductor (MOS) CCDs
\citep{2001AAturner}\ and the third one uses pn CCDs  \citep{2001AAstruder}\
for data recording. Also, there are two Reflection Grating Spectrometers
 \citep{2001AAdenherder}\ located behind two of the telescopes.
V2491 Cyg was observed (pointed archival observation, OBSID=0552270501) for an
exposure of 39.2 ks between
2008 May 20 UT 14:03:53 and 2008 May 21 UT 00:59:28 ($\sim$40 d after outburst). The second 
pointed archival observation (OBSID=0552270601) was obtained
between 2008 May 30 08:21:01 and 2008 May 30 16:40:40 for an \XMM\ exposure of 29.8 ks ($\sim$50 d after outburst). 
Both observations used
all the EPIC cameras with different modes, but, in this study  we utilize the
RGS data to obtain  the high resolution spectra for our analysis purpose.

The RGS observations were carried out using the standard "High Event Rate with SES" spectroscopy mode for readout.
For the analysis, Science Analysis Software (SAS)
version 14.0.0 was utilized and we reprocessed the data using the XMM-SAS routine {\tt rgsproc}.
Source and background counts for the RGS data were extracted using the standard spatial and energy filters;
for the source position,
which defines the spatial extraction regions as well as the wavelength zero point.
{\tt rgsproc} allows the user to restrict the processing to an enumerated subset of exposures within the
observation and an enumerated set of reflection orders. From the first stage to the fifth stage 
it performs basic calibrations
on the events in separate CCD event lists and then gathers them in combined event list; next
does source-specific aspect-drift correction and defines the channel grids for events and exposures
allowing filtering of data (to correct or to remove what is unusable, e.g., bad-pixel correction).
The fourth and fifth stages produce spectra and generate response matrices for the designated primary source. 
The sixth stage produces light curves. Following this,
we derived spectra for different orders of RGS1 and RGS2 and generated light curves. In addition,
we used the SAS task  {\tt rgsfluxer} to obtain fluxed spectra between 5-38 \AA\ with 3400 bins as it is required for the 
SPEX software.  
In addition, we  used the event files and determined times of
low background from the count rate on CCD 9 (which is closest to the optical axis).
The final exposure times and net count rates
showed that there were no sporadic high background events in our data and since the
nova was  bright in both observations, we did not include any corrections for flares.
Figure 1 shows the
light curves of the first and second observations (RGS1 data) with a bin time of 100 s. 
Following the analysis by \citet{2011ApJness}\ and 
\citet{2012AApinto}, we selected the three different count rate regions in the light curve of the first observation
as labeled in the figure (i.e., region1, region2, region3) to investigate the nature and effect of the variation seen in the 
light curve on the produced spectra
from the given regions. The second observation did not show any major variation, thus the entire event file was used for the analysis. 
Next, we extracted the events from the event files by filtering on time for the first observation (region1-3)
and then re-run {\tt rgsproc} and {\tt rgsfluxer} to produce appropriate response matrices and fluxed spectra.  

\begin{figure*}
\centerline{
\includegraphics[width=4.6in,height=2.8in,angle=0]{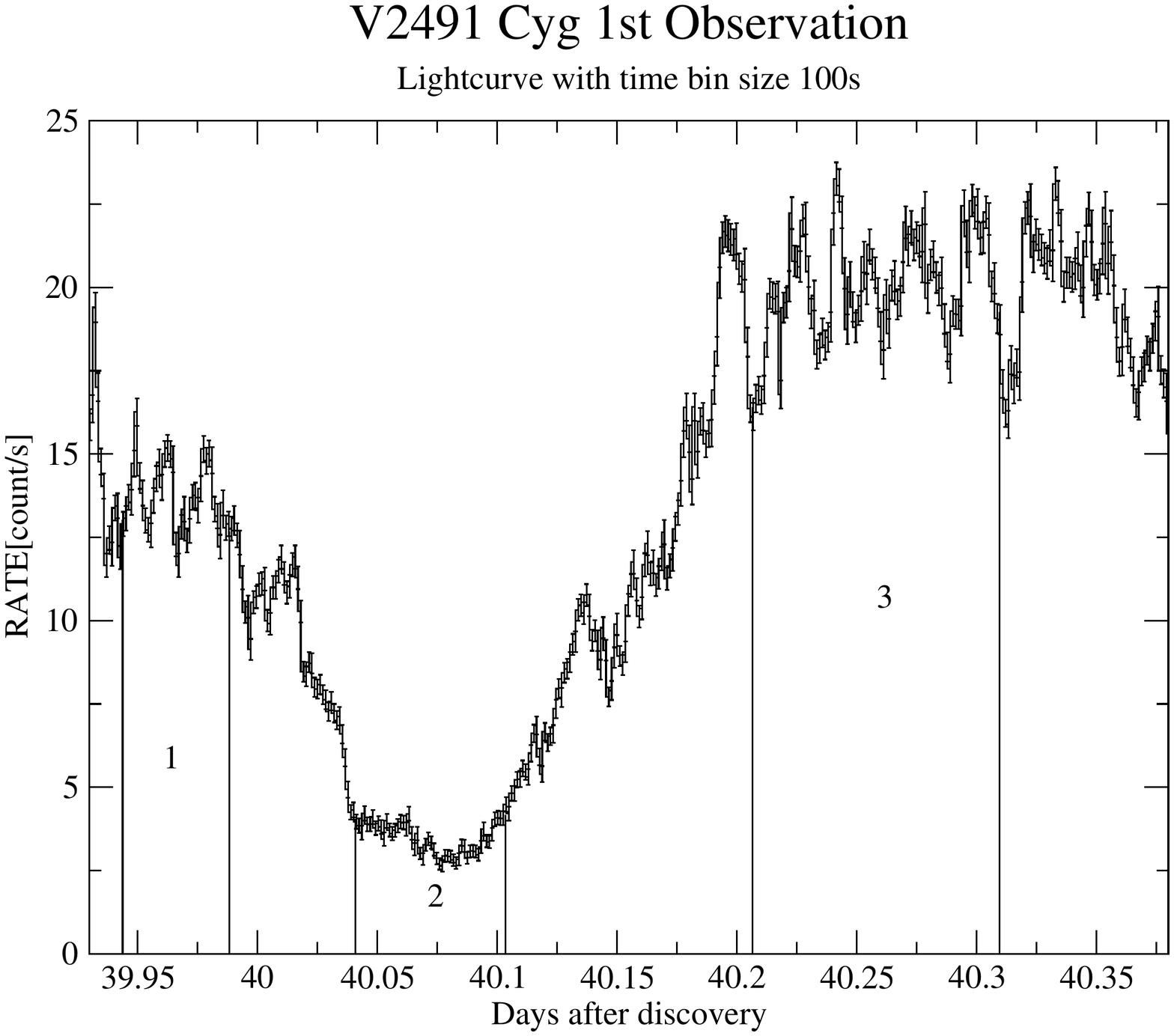}
\includegraphics[width=3.5in,height=2.8in,angle=0]{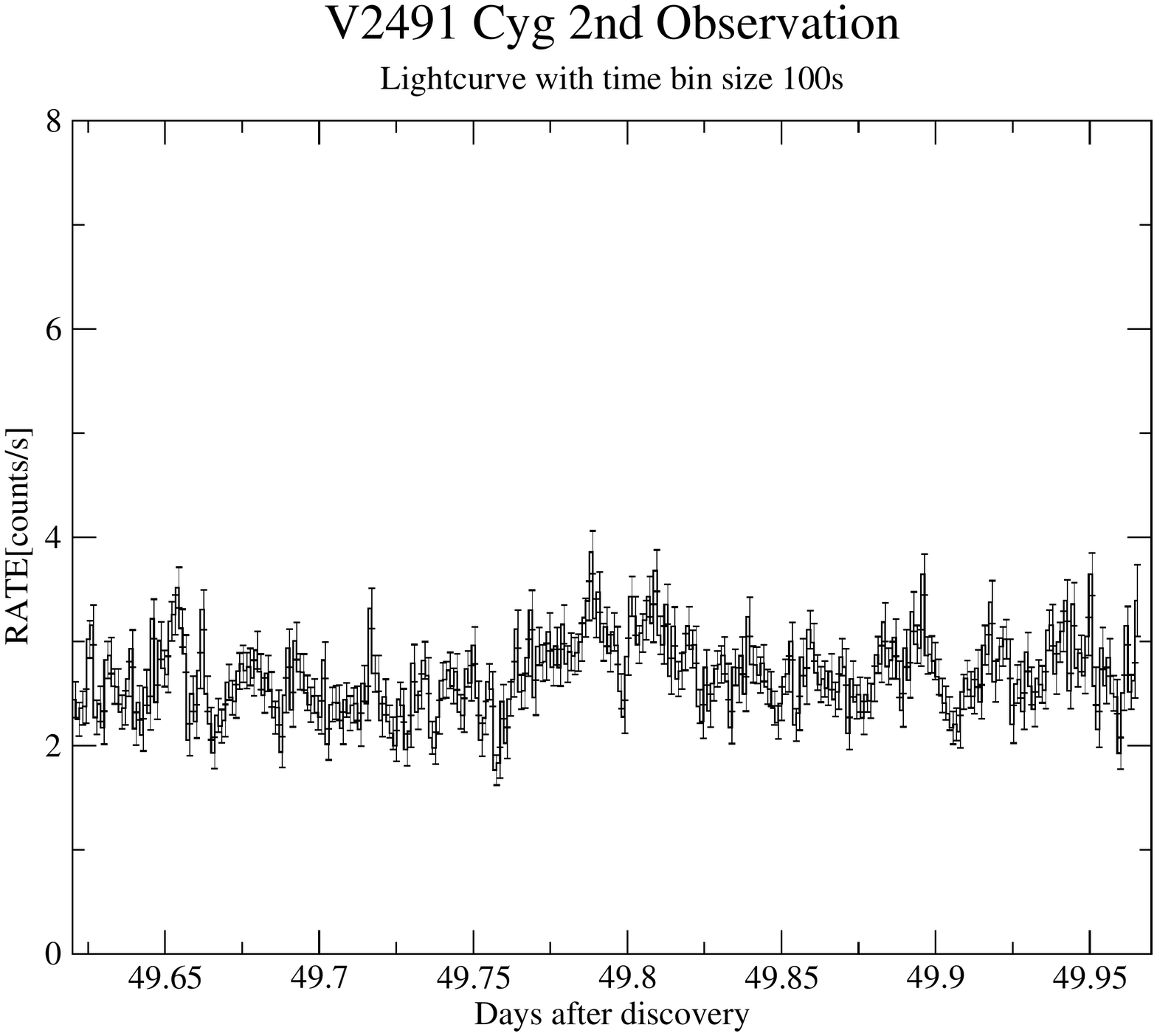}}
\caption{The \XMM\ RGS1 light curve (LC) of V2491 Cyg. First observation (40 d after outburst) 
is on the left and the second observation is on the right (50 d after outburst). The three different regions of
count rates are labeled as 1, 2, and 3 on the LC of the first observation as assumed in the spectral analysis.}
\end{figure*}


\section{Analysis and Results}
\subsection{Photoionized and collisionally ionized absorber models}
\label{subsect:models}

SPEX is a software package \citep[version 2.05.04;][]{1996kaastra}\ optimized for the analysis and interpretation of
high-resolution cosmic X-ray spectra. The software is especially suited for
fitting spectra obtained by current X-ray observatories like \XMM, $Chandra$,
and $Suzaku$. It is maintained by SRON (HEA Division, Netherlands Institute for
Space Research).

We obtained a similar model description to \citet{2012AApinto}\ to fit both sets of the nova data (first and second
observation) in order to
compare the different ionized absorber components in the system. There are two main models in SPEX 
that can be used  to model ionized intrinsic absorption; one of them
is the {\tt xabs}  \citep[used by][]{2012AApinto}\ and the other is the {\tt Hot} model used in this study.
The {\tt xabs} model of SPEX calculates the warm, photoionized
absorption by a thin slab composed of different ions located between the ionizing source and the
observer. The {\tt xabs} model assumes a small angle subtended by the slab at the source so that
only absorption and scattering
out of the line-of-sight by the slab are considered. The processes taken into account are the
continuum and the line absorption by the ions and scattering out of
the line-of-sight by the free electrons in the slab. The transmission, T, of the slab is
calculated as $T = \exp[-\tau_{\rm c} -\tau_{\rm l} -\tau_{\rm e}]$, here $\tau_{\rm c}$ and
$\tau_{\rm l}$ are the total continuum and line optical depth, respectively, and $\tau_{\rm e}$
is the electron scattering optical depth.
For $\tau_{\rm e}$ the classical Thomson approximation is valid
below 10 keV (longer than 1.24 \AA\ ).
Most continuum opacities are taken from \citet{1995AASverner}, while line
opacities and wavelengths for most ions are taken from  \citet{1996ADNDTverner}\ and abundances are calculated
using Lodders $\&$ Palme (2009) (see the details and
additional references in the SPEX user's manual). $\tau_{\rm l}$ is a function of v and $\sigma _{\rm v}$
which are free parameters of {\tt xabs}. v is the average systematic velocity shift of the absorber, in km s$^{-1}$.
Negative and positive values of v correspond to blue and red shifts, respectively. $\sigma _{\rm v}$ is
the turbulent velocity broadening of the absorber in km s$^{-1}$, defined as
$\sigma_{\rm total}^{2} = \sigma_{\rm v }^{2} + \sigma_{\rm thermal}^{2}$, where
$\sigma_{\rm total}$ is the total width of a line and $\sigma_{\rm thermal}$ the thermal contribution.
The relative column densities of
the ions are calculated through a photoionization model introducing two free parameters:
N$_{\rm H^{xabs}}$ and $\xi$. N$_{\rm H^{xabs}}$ is the equivalent hydrogen column density of the
ionized absorber in units of atoms cm$^{-2}$. $\xi$ is the ionization parameter of the absorber
defined as $\xi = L /n_{\rm e} ~ r^{2}$, where L is the luminosity of the ionizing source,
n$_{\rm e}$ the density of the plasma and r the distance between the slab and the
ionizing source. $\xi$ is expressed in units of erg cm s$^{-1}$.
Codes such as XSTAR\ \citep{2001ApJSkallman}\ or CLOUDY\ \citep{2003ARAAferland}\ are used for the
broad-band ionizing continuum from infrared to hard X-rays where the ionic column densities of a
photoionized slab is  precalculated for different values of $\xi$
and used in the fitting procedure with SPEX. During the fitting process, SPEX reads in the grid of precalculated ionic column 
densities and finds the best 
set and the best-fit values for $N{\rm _H^{xabs}}$ and $\xi$. 

In order to compare with the analysis using {\tt xabs} models, here, we focus mainly on the
{\tt Hot} model within SPEX that assumes
a collisional ionization (equilibrium) absorber model instead of the photoionized absorber model.
This model calculates the transmission of a plasma in collisional equilibrium with
cosmic abundances. For a given temperature and abundances, the model calculates the ionization balance
and then determines all ionic column densities by scaling the prescribed hydrogen column density.
Using these column densities, the transmission of the plasma is calculated by multiplying the
transmission of the individual ions. The transmission assumes both continuum and line opacity.
Most continuum opacities are taken from \citet{1995AASverner}, while line
opacities and wavelengths for most ions are taken from  \citet{1996ADNDTverner}\ and abundances are
calculated using Lodders $\&$ Palme (2009).
This model will mimic neutral plasma transmission at about 0.0005 keV (6000 K) which we utilize to model the
gas component of the interstellar absorption. It, also, has similar
free parameters to {\tt xabs} model like velocity shift v and turbulent velocity broadening $\sigma_{\rm v}$ together with
equivalent hydrogen column density N$_{\rm H^{Hot}}$ and kT, temperature of the absorber in keV.

\subsection{\XMM\ RGS Spectrum of V2491 Cyg }

As outlined in section [2.0], we analyzed the data using Science Analysis Software (SAS)
version 14.0.0 and produced RGS1-2 spectra and relevant response matrices using {\tt rgsproc} and {\tt rgsfluxer}
for different spectral orders.
Some spectral analysis checks were performed using XSPEC \citep{1996arnaud} version 12.9.0, and the
main analyses were conducted with the SPEX software package
\citep{1996kaastra} version 2.05.04\ .  In order to perform the analysis with SPEX, RGS1 and RGS2 first order spectra were 
combined by using tasks {\tt rgsfluxcombine} and  {\tt rgsfmat} within SPEX. 
The fluxed spectra were created between 5-38 \AA\ with 3400 bins.
In general, the spectra between 7-38 \AA\ (0.3-1.8 keV) were used for the fitting process and the data were binned by a factor of 
4 between 7-11 \AA\, and by a factor 2 for 11-38 \AA\ . 
The detailed description of the emission and absorption lines and edges, can be found in  \citet{2011ApJness}\ and
\citet{2012AApinto}.
  
Our main aim is to study the complex absorption properties in the high resolution X-ray spectra
of classical and recurrent novae and our target, in this work, is V2491 Cyg. In order to compare with the 
absorption modeling used in \citet{2012AApinto}\ which used the photoionized absorber model {\tt xabs},  
we used the absorption model {\tt Hot} which is a collisionally ionized (in equilibrium) absorber model in the
SPEX software package. Since a comparative study with the {\tt xabs} models would require that the data are analyzed
and fitted in the same manner, we followed the analysis steps and used the same modeling in \citet{2012AApinto}\
except for the photoionized absorber models. We utilized a composite model that includes a blackbody model  {\tt  Bb} for the continuum
and a plasma emission model in collisional equilibrium  {\tt Cie} for the excess in the harder X-ray band together with
two {\tt Hot} absorber models, one {\tt Amol} model for the dust absorption \citep{2010AApinto} in the line of sight and finally another
{\tt Hot} model with a very low temperature ($\sim$ 1eV) that would mimic the cold gas absorption in the interstellar medium. 
We followed a similar fitting procedure to  \citet{2012AApinto}. First the harder X-ray band data (7-14.4 \AA) was fitted
with the {\tt Cie} model together with the cold {\tt Hot} ({\tt Hot-ISM}) absorber model for the gas in the interstellar medium
to determine the abundances separately from the other absorber models using the fluxed 
spectrum of region3 and the second observation. 
An additional blackbody model {\tt Bb} was added in the fitting process for the 
second observation to account for the long wavelength excess. If the additional blackbody model was not included, the 
 \rchisq\ of the fit was 2.15 and not acceptable.
Figure 2 shows the fitted  models to the two data sets.
We find that the spectral
parameters of this blackbody emission are different than the main blackbody model parameters of the
stellar remnant presented in Table 1 (\& 2). The norm is (2.0-4.0)$\times$10$^{15}$ cm$^{2}$, temperature
is (120-131) eV and the luminosity is 6.0$\times$10$^{35}$ erg s$^{-1}$. However, inclusion of this component
in the fitting procedure in the total RGS band 7-38 \AA\ does not affect the spectral parameters in Table 1 (\& 2), 
particularly the main stellar remnant emission, 
and it only improves the fits at about 93$\%$ confidence level.
The spectral parameters of this new blackbody component remain similar in the 90$\%$-95$\%$ confidence level when 
fitted in the 7-14 \AA\ or 7-38 \AA\
energy range. We elaborate on this component in the discussion section.
 
The Ne and Mg abundances from these fits above (i.e., using region3 and second observation spectra) were fixed 
together with the $v_{\rm mic}$ parameter for the fits using fluxed spectra of regions 1 and 2. Note that the free fits yield very similar 
abundances of Ne and Mg for the second observation.
Following this, first the fluxed spectrum of region3 was fitted with the composite model
{\tt Hot-ISM$\times$(Cie+(Hot-1$\times$Hot-2)$\times$(Bb))} 
and once the fit converged the multiplicative model {\tt Amol} 
was added and a final fit was performed {\tt Amol$\times$Hot-ISM$\times$(Cie+(Hot-1$\times$Hot-2)$\times$(Bb))}. 
Next, the elemental abundances derived for the {\tt Hot} absorber models using the fit to the fluxed spectrum of region3
were fixed for all the fits to the fluxed spectra of the regions 1, 2 and the second observation 
assuming that the abundances in the ejecta/wind will stay the same in the short duration of the X-ray observation, and that
the second observation is only 10 days after the first observation.
The fitted fluxed spectra of V2491 Cyg can be found in Figures 3-6 where Figure 3-5 are for regions 1, 2, and 3 as labeled 
in Figure 1 (left panel) for the first observation. Figure 6 is the spectral fit to the fluxed spectrum of the second observation.
The resulting spectral parameters from fits with the composite
model using the collisionally ionized (in equilibrium) hot absorber models are listed in Table 1. All abundances are given in Solar units.
The error ranges of the spectral parameters are given at the 90$\%$ confidence level for a single parameter ($\Delta\chi^2$=2.71). 
The degrees of freedom are taken in accordance with the  \citet{2012AApinto}\ fits for comparison. However, we assumed
abundances in Solar units which takes H as the reference element instead of using oxygen as the reference. 

\begin{figure*}
\includegraphics[width=3.6in,height=2.6in,angle=0]{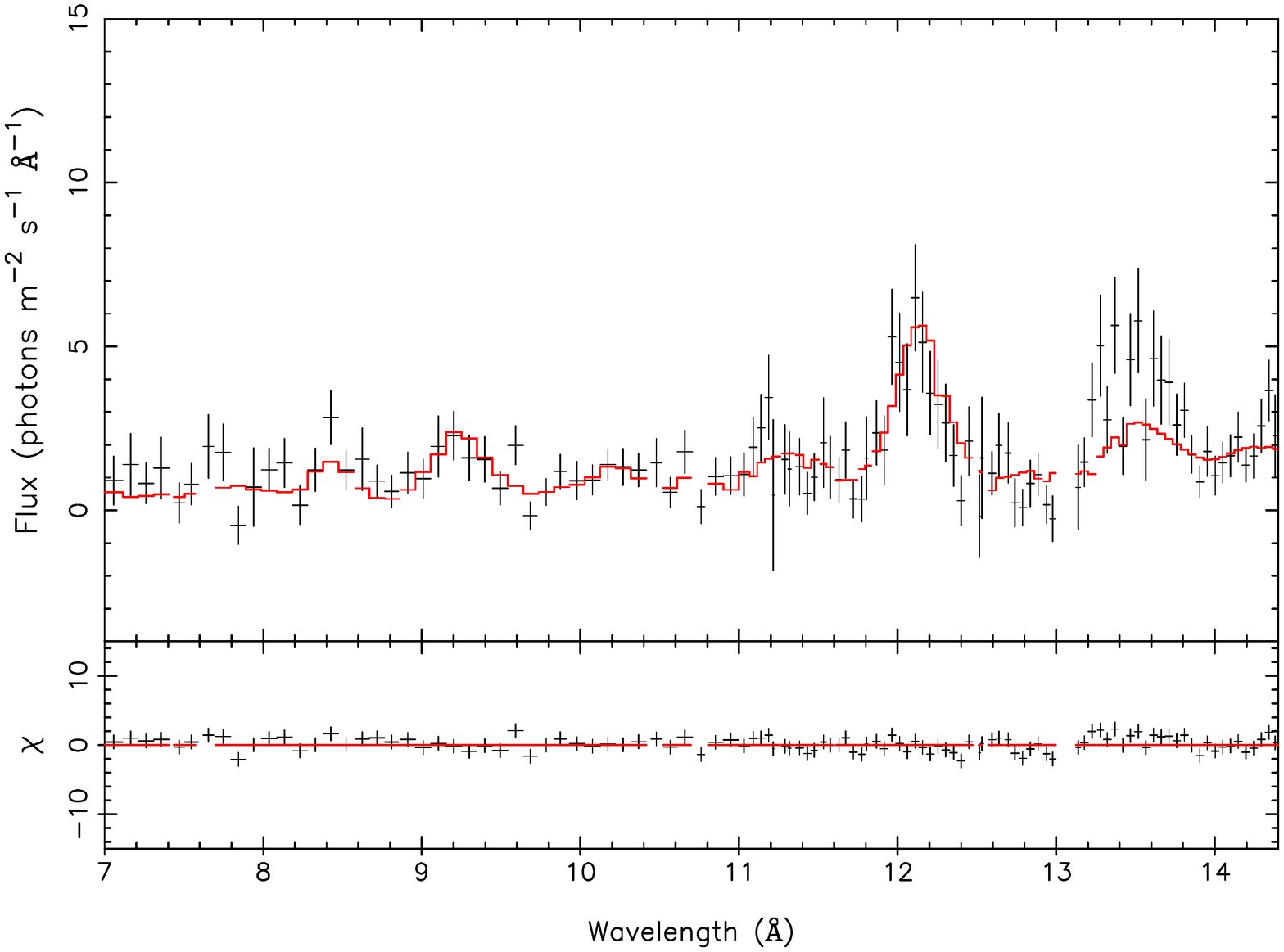} 
\includegraphics[width=3.6in,height=2.6in,angle=0]{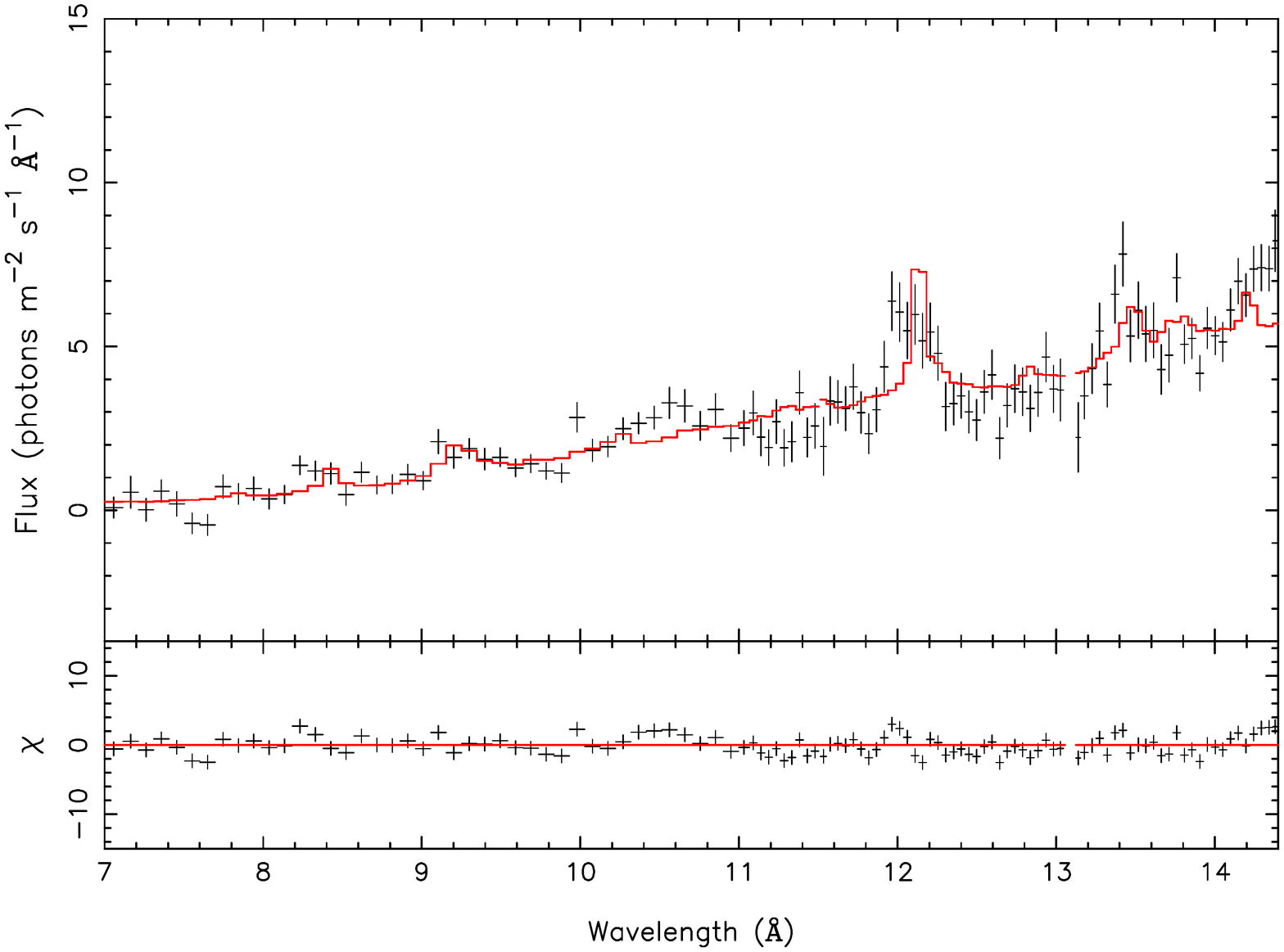}
\caption{The RGS1+2 fluxed spectrum of V2491 Cyg fitted in the range 7-14.4 \AA\ using the plasma emission model
{\tt Cie} and the cold {\tt Hot-ISM} absorption model. A blackbody model of emission was also used 
as an additive model for the second observation spectrum. 
The left hand and right hand panels show the
fit to the fluxed spectrum of region3 and second observation data, respectively.  }
\end{figure*}

We find that our fits yield  $\chi_\nu^{2}$=1.8-2.9 (for about 1440 degrees of freedom) using
the collisionally ionized hot intrinsic absorber models for the classical nova V2491 Cyg. 
Given the $\chi_\nu^{2}$
values, the collisionally ionized absorber models yield considerably better fits
to these data sets of V2491 Cyg than the photoionized absorber fits in  \citet{2012AApinto}\ which have  $\chi_\nu^{2}$ in a range
2.3-7.0 for about 1470 degrees of freedom.
Other modeling using expanding atmosphere models \citep{2010ANvanrossum} yielded very high values of $\chi_\nu^{2}$ not given 
in the paper and the NLTE static atmosphere fits in \citet{2011ApJness} had a  $\chi_\nu^{2}$ range of 
2.2-21.8 for about 5600 degrees of freedom.
We note that the comparison with the  \citet{2012AApinto}\ results using the photoionized absorbers
indicates that particularly, for the fits to the region1 and region3 spectra where the count rates are the highest,
inclusion of the collisionally ionized hot absorbers improves the fits 
to a good extent, 
the improvements for the low count rate region2 and the second observation data with the lowest count rate are rather 
less obvious. 
We believe this is because the high count rate data show the absorption and emission features (along with a strong continuum)
more prominently allowing for better fitting of the detailed models using the high spectral resolution data.
Such high spectral resolution data with low count rates have increased statistical errors with the detailed spectral
features more smeared out or less visible (with weaker continuum) which does not allow for the fit quality to differentiate between
several models appropriately. However, in all four spectra that we have fitted, we have better \rchisq\
values in comparison with the intrinsic photoionized absorber fits in \citet{2012AApinto}. 
An FTEST (within XSPEC) can be performed to test the significance of improvement of these fits with respect to the photoionized
absorber fits in  \citet{2012AApinto}. Comparison of the \chisq\ and degrees of freedom of each fit for regions 1-3, the FTEST  probability
ranges between 3$\times$10$^{-26}$-6$\times$10$^{-270}$ which is the value of 1-(confidence level). Thus, the improvement is much better than 10$\sigma$
where 3$\sigma$ significance would have FTEST probability of $\sim$\ 0.003\ . However, the FTEST for the fits
to the lowest count rate data (the second observation) gives a probability of 0.4 (about 60\% confidence level) 
which yield only around 1$\sigma$ improvement for the collisionally ionized absorber fits over the photoionized absorber fits. 

In addition, we tested the possibility of having a third absorber by adding another {\tt Hot} model to the fits
in Table 1 and tested the significance of this component by applying an FTEST. 
We find that the fits to the data of region 1, 2, and the second observation yield improvement of \chisq\ values at a
confidence level less than 30\%. However, FTEST for the fit of the region3 spectrum resulted in a probability
of 5$\times$10$^{-8}$ yielding an improvement over 5$\sigma$. 

Our original idea, as described in the Introduction section, was a composite model where there is complex absorption of interstellar, 
photospheric, and of collisionally-ionized and/or photoionized gas originating from the moving material in
the line of sight from a nova wind or ejecta. As a result, our goal was to model 
the absorption components detected in the high resolution spectra independently from the continuum
model (i.e., mainly stellar remnant emission) 
and compare the photoionized warm and collisionally ionized hot absorber gas. Thus, next, we employed the {\tt xabs} model
for the third absorber in V2491 Cygni ( {\tt Amol$\times$Hot-ISM$\times$(Cie+(Hot-1$\times$Hot-2$\times$xabs)$\times$(Bb))}). 
This composite model implies that there is some absorption from photoionized gas along with the
main absorbers that are collisionally ionized. 
For this additional model, we used a grid of ionic column densities for {\sc xabs} calculated utilizing the CLOUDY software
assuming an ionizing continuum of blackbody radiation at 65 eV $(7.8\times10^5 K)$ chosen inaccordance with our analysis.
The resulting fits show that all the \rchisq\ values except the region2 fit 
(FTEST probability is 0.005
yielding an improvement at 99.5$\%$ confidence level) are improved
over  10$\sigma$ significance (using  FTESTs the probability is 3.2$\times$10$^{-12}$-6.9$\times$10$^{-47}$). 
These improved fits are given in Table 2. We note here that the spectral parameters (in Table 1 \& 2) of the
main stellar remnant emission ({\tt Bb}), plasma emission {\tt Cie}, the interstellar dust {\tt Amol} and {\tt Hot-ISM} gas absorption
remain consistent within 90$\%$-95$\%$ confidence level errors. 
  
Finally, we note that
 \citet{2012AApinto}\
have attempted to use stellar atmosphere models instead of blackbody models using particular CNO abundances along with
the complex absorption model they have used. They report that such models only resulted in higher  $\chi_\nu^{2}$ values and
have not improved the fits at all.
  
\begin{figure*}
\includegraphics[width=6.9in,height=3.8in,angle=0]{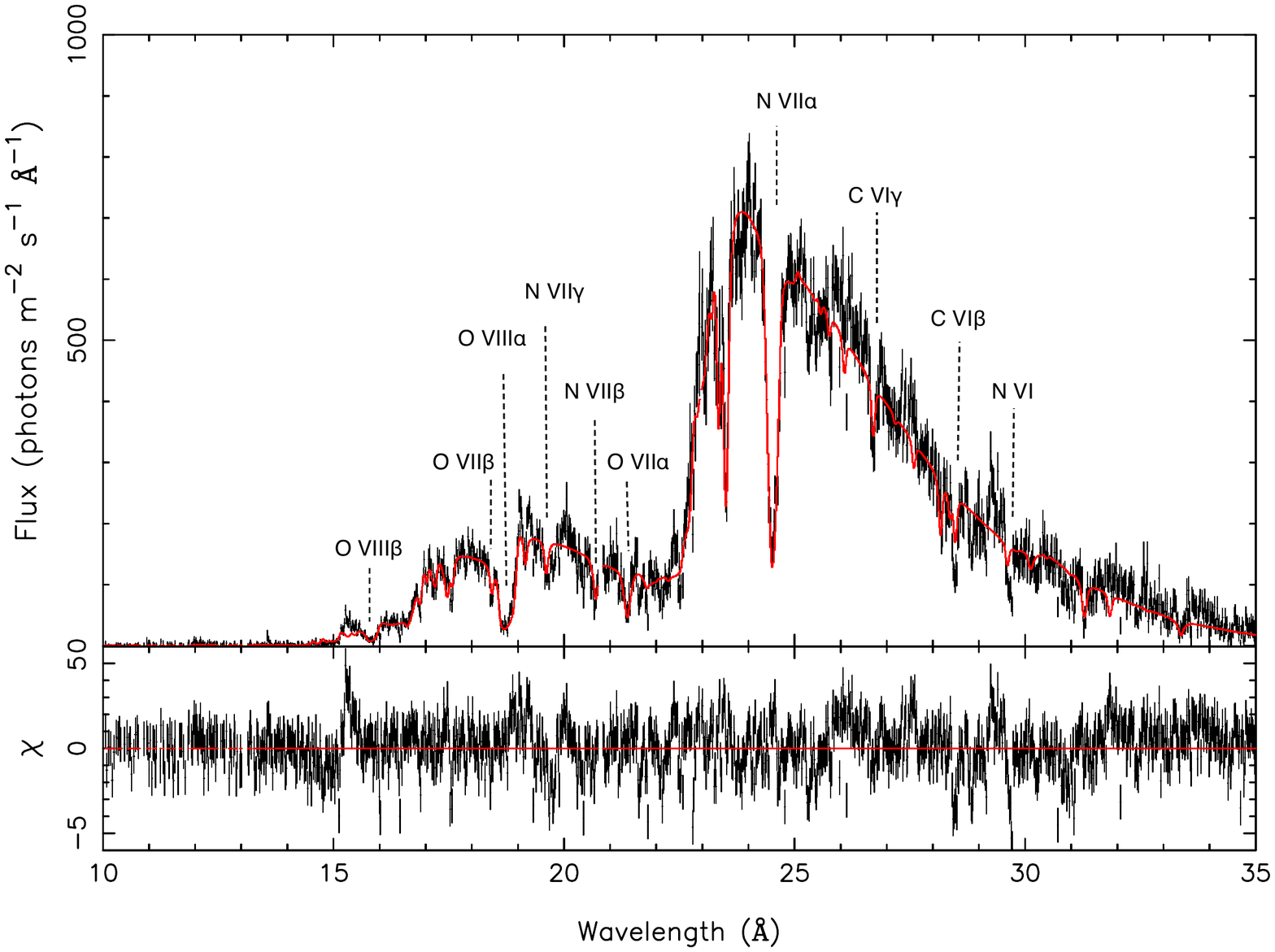}
\caption{The RGS1+2 fluxed spectrum of V2491 Cyg using region1 of the LC (first observation)  fitted with the composite model
described in the text using the {\tt Hot} model 
in SPEX (see Table 1). Certain detected blueshifted absorption lines are labeled with dashed lines.}
\end{figure*}

\begin{figure*}
\includegraphics[width=6.9in,height=3.8in,angle=0]{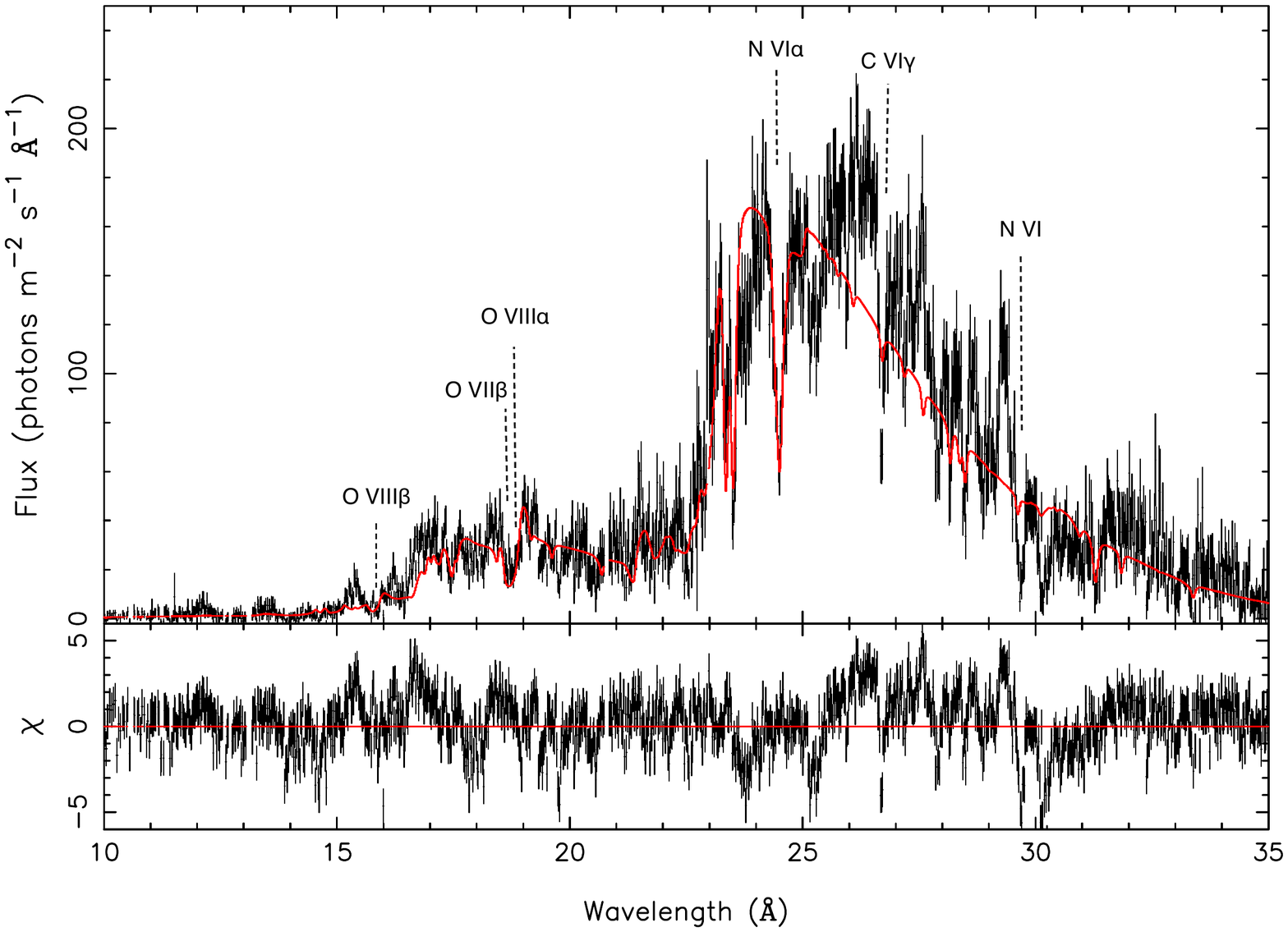}
\caption{The RGS1+2 fluxed spectrum of V2491 Cyg using region2 of the LC (first observation) fitted with the composite model
described in the text using the {\tt Hot} model in SPEX (see Table 1).
Certain detected blueshifted absorption lines are labeled with dashed lines.}
\end{figure*}

\begin{figure*}
\includegraphics[width=6.9in,height=3.8in,angle=0]{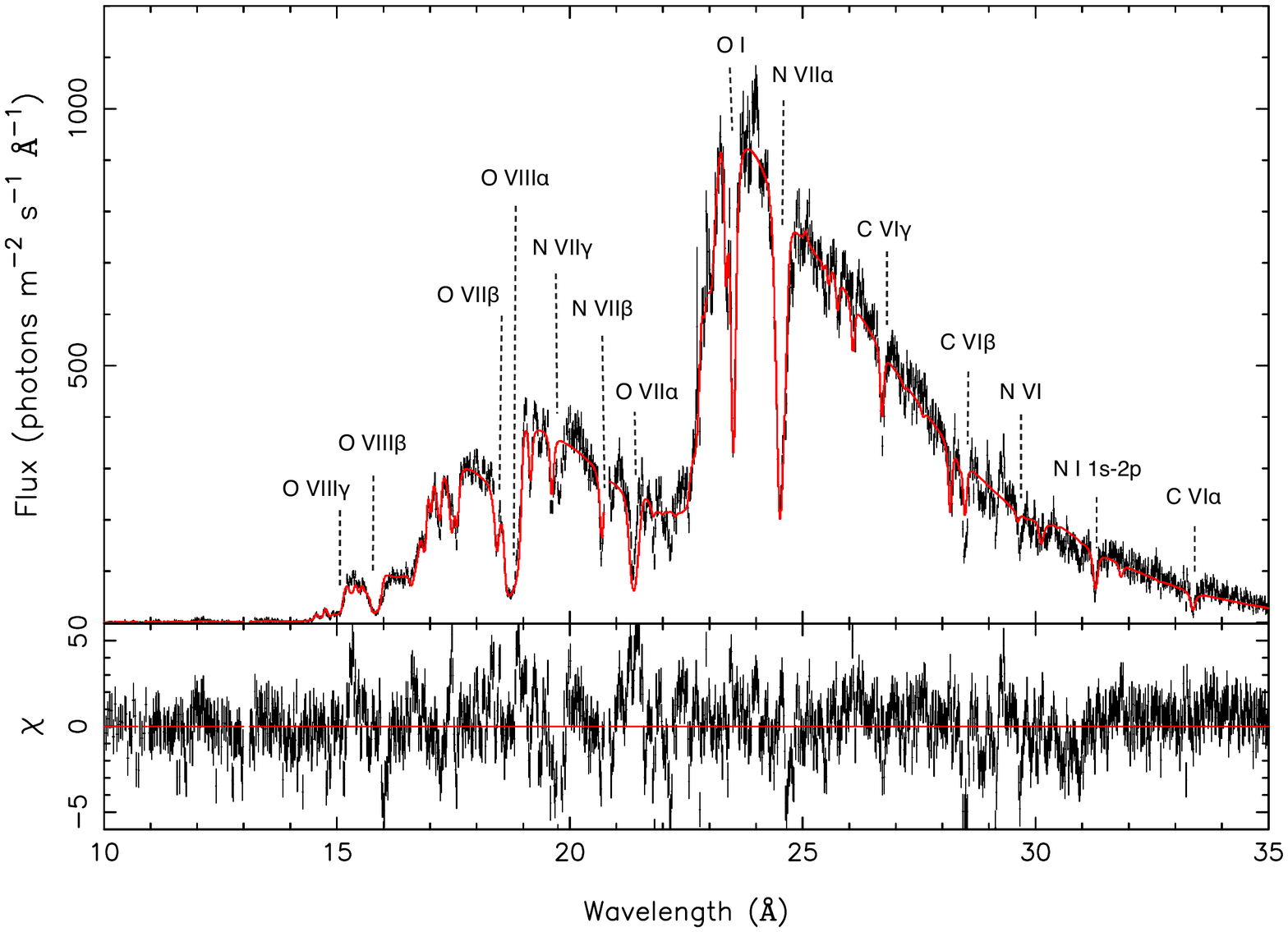}
\caption{The RGS1+2 fluxed spectrum of V2491 Cyg using region3 of the LC (first observation) fitted with the composite model
described in the text using the {\tt Hot} model
in SPEX (see Table 1). Certain detected blueshifted absorption lines are labeled with dashed lines.}
\end{figure*}

\begin{figure*}
\includegraphics[width=6.9in,height=3.8in,angle=0]{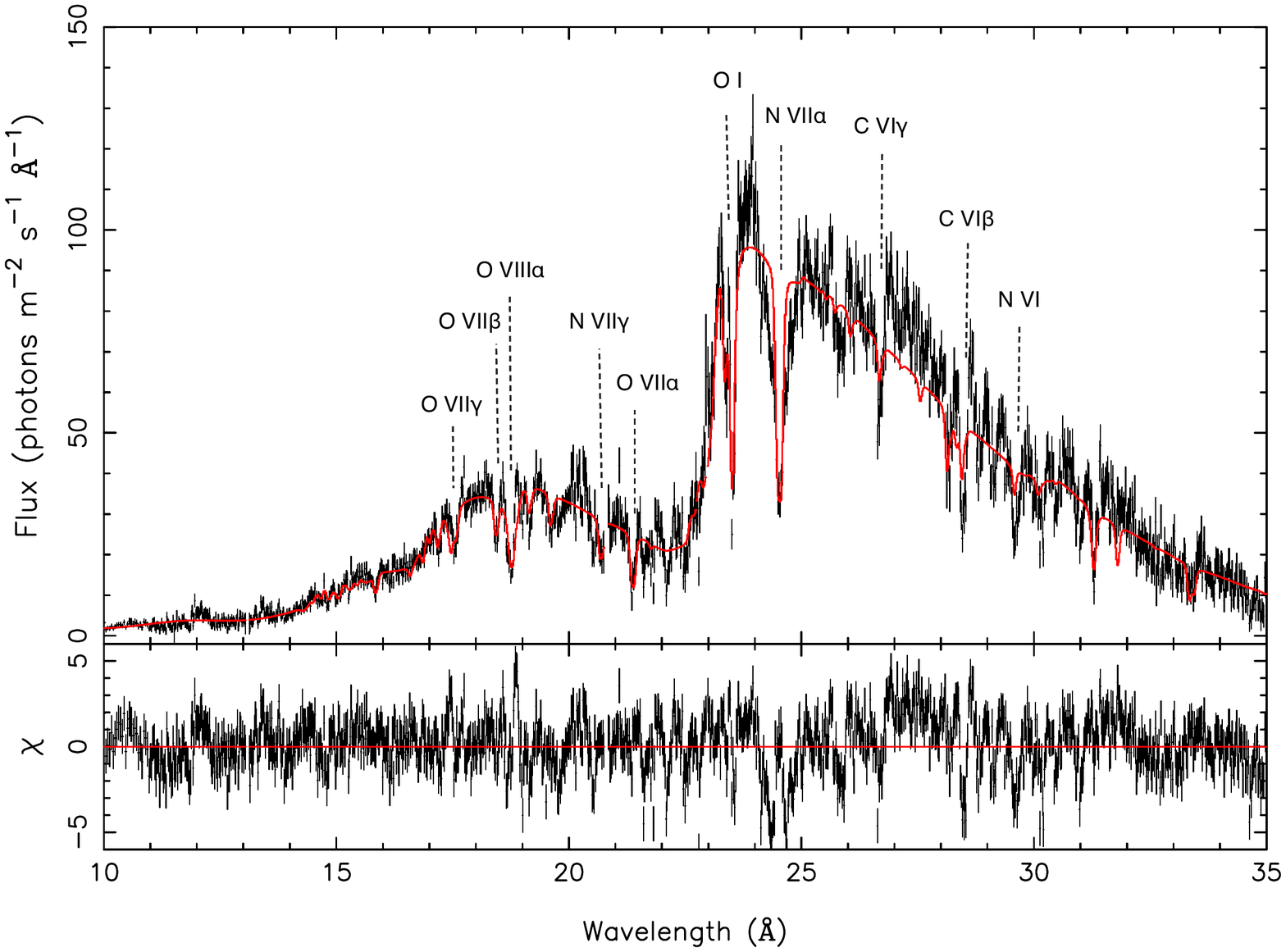}
\caption{The RGS1+2 fluxed spectrum of V2491 Cyg obtained from the second observation fitted with the composite model
described in the text using the {\tt Hot} model
in SPEX (see Table 1).  Certain detected blueshifted absorption lines are labeled with dashed lines.}
\end{figure*}

\begin{table*}
  \caption{The spectral parameters obtained from the composite model fits including two {\tt Hot} absorber
models to the RGS spectrum of V2491 Cyg (errors are calculated at the 90\% confidence level.)}\label{1}
  \centering 
  \begin{tabular}{l|l|c|c|c|c}
\hline
   Model & Parameters & Region1  & Region2& Region3& $2^{nd}$ Obs.\\ \hline\hline
   
   Blackbody &Norm ($10^{20} \rm cm^2$) &$1.53^{+0.39}_{-0.25}$&$0.23^{+0.86}_{-0.04}$&$0.40^{+0.47}_{-0.21}$&$0.02^{+0.05}_{-0.01}$\\
                     &Temperature (eV)            &$68.5^{+0.9}_{-0.9}$&$64.5^{+0.6}_{-3.6}$& $88.4^{+2.7}_{-1.8}$ &$75.7^{+9.1}_{-6.5}$\\
                     &Luminosity($10^{39}$ erg s$^{-1}$)&2.18&0.23&1.92&0.05\\\hline

   CIE     & Norm ($10^{58} \rm cm^{-3}$)&$0.40^{+0.11}_{-0.09}$&$0.72^{+0.79}_{-0.32}$&$0.42^{+0.13}_{-0.05}$&$0.27^{+0.06}_{-0.1}$\\
              &Temperature (keV)         &$0.20^{+0.01}_{-0.01}$&$0.16^{+0.03}_{-0.02}$&$0.20^{+0.01}_{-0.01}$&$0.30^{+0.13}_{-0.03}$\\
              & $v_{mic}$ (km s$^{-1}$)           &2791&2791& $2791^{+333}_{-288}$&$25068^{+1383}_{-951}$\\
              & Ne Abundance              &2.45&2.45& $2.45^{+1.36}_{-0.92}$&$2.4^{+0.9}_{-0.9}$\\
              & Mg Abundance              &1.2&1.2& $1.2^{+1.0}_{-0.6}$&$1.98^{+0.70}_{-0.70}$\\
              &Luminosity($10^{36}$ erg s$^{-1}$)&1.31&1.78&1.38&0.77\\ \hline

   HOT-1 & $N_H$ ($10^{24} \rm cm^{-2}$)       &$0.40^{+0.03}_{-0.03}$&$0.09^{+0.03}_{-0.01}$& $0.07^{+0.01}_{-0.01}$&$0.19^{+0.10}_{-0.02}$\\
              &Temperature (keV)                      &$2.63^{+0.96}_{-0.56}$&$1.77^{+0.17}_{-0.12}$& $1.06^{+0.03}_{-0.03}$&$1.38^{+0.58}_{-0.12}$\\
              & $\sigma_v$ rms velocity (km s$^{-1}$) &$872^{+25}_{-24}$&$827^{+61}_{-86}$& $879^{+12}_{-12}$&$114^{+18}_{-15}$\\
              & velocity shift (km s$^{-1}$)                   &$-3186^{+51}_{-50}$&$-3699^{+113}_{-173}$& $-3128^{+30}_{-31}$&$-2636^{+47}_{-45}$\\ \hline

   HOT-2 & $N_H$ ($10^{24} \rm cm^{-2}$)       &$0.86^{+0.03}_{-0.03}$&$0.18^{+0.01}_{-0.02}$&$1.80^{+0.02}_{-0.02}$&$0.51^{+0.02}_{-0.05}$\\
                 &Temperature (keV)                      &$0.81^{+0.03}_{-0.03}$&$0.58^{+0.07}_{-0.11}$& $0.99^{+0.03}_{-0.03}$&$0.82^{+0.05}_{-0.45}$\\
              & $\sigma_v$ rms velocity (km s$^{-1}$) &$54^{+8}_{-9}$&$<9$& $56^{+4}_{-5}$&$60^{+7}_{-8}$\\
              & velocity shift (km s$^{-1}$)                     &$-3180^{+71}_{-58}$&$-3038^{+6}_{-183}$& $-3194^{+30}_{-31}$&$-3550^{+36}_{-33}$\\\hline

  Abundances & C Abundance            &0.43&0.43& $0.43^{+0.07}_{-0.07}$&0.43\\
                       & N Abundance            &5.2&5.9& $5.9^{+0.5}_{-0.9}$&5.9\\
                       & O Abundance            &37.9&37.9& $37.9^{+5.8}_{-10.0}$&37.9\\
                       & Si Abundance            & 0.02& 0.02& $<0.02$& 0.02\\
                       & S Abundance             &0.02&0.02& $0.02^{+0.01}_{-0.01}$&0.02\\
                       & Ar Abundance            &1.6&1.6& $1.6^{+3.9}_{-1.2}$&1.6\\
                       & Ca Abundance           & 0.01& 0.01& $<0.01$& 0.01\\
                       & Fe Abundance           &8.9&8.9& $8.9^{+3.8}_{-2.1}$&8.9\\\hline

   HOT-ISM & $N_H$ ($10^{21} \rm cm^{-2}$)       &$3.89^{+0.06}_{-0.06}$&$3.84^{+0.06}_{-0.19}$& $3.26^{+0.05}_{-0.09}$&$2.4^{+0.3}_{-0.2}$\\
                   &Temperature (eV)                    &$1.11^{+0.03}_{-0.04}$&$1.20^{+0.02}_{-0.04}$& $1.01^{+0.02}_{-0.02}$&$0.8^{+0.3}_{-0.3}$\\
                   & N Abundance                          &1.16&1.16 & $1.16^{+0.12}_{-0.12}$&1.16\\
                   & O Abundance                          &1.75&1.75 & $1.75^{+0.04}_{-0.02}$&1.75\\
                   & Fe Abundance                         &0.95&0.95& $0.95^{+0.13}_{-0.12}$&0.95\\\hline

   AMOL (Dust) &$O_2$ ($10^{17} \rm cm^{-2}$)            & $0.34^{+0.13}_{-0.13}$& $2.3^{+0.2}_{-0.2}$&$<0.1$&$1.3^{+0.1}_{-0.1}$\\
                           &$H_2O (ice)$ ($10^{17} \rm cm^{-2}$) & $5.5^{+0.3}_{-0.3}$&$7.8^{+0.5}_{-0.5}$ &4.77$^{+0.13}_{-0.13}$ &$7.2^{+0.2}_{-0.2}$\\
                           &$CO$ ($10^{17} \rm cm^{-2}$)             &$ 1.28^{+0.13}_{-0.13}$&$0.27^{+0.2}_{-0.2}$ & $0.1^{+0.07}_{-0.06}$&$<0.1$\\\hline
                           &\rchisq         &1.80 &2.4 &2.86&2.3 \\
                           &  (d.o.f.)                &(1415)&(1439) &(1440)&(1465) \\\hline

\end{tabular}
\end{table*}


\begin{table*}
 \caption{The spectral parameters obtained from the composite model fits including two {\tt Hot} absorber
and one  {\tt xabs} models to the RGS spectra of V2491 Cyg (errors are calculated at the 90\% confidence level.)}\label{2}
  \centering
  \begin{tabular}{l|l|c|c|c|c}
\hline
   Model & Parameters & Region1  & Region2& Region3& $2^{nd}$ Obs.\\ \hline\hline

   Blackbody &Norm ($10^{20} \rm cm^2$) &$0.99^{+0.06}_{-0.06}$&$0.46^{+0.05}_{-0.05}$&$0.40^{+0.01}_{-0.01}$&$0.02^{+0.01}_{-0.01}$\\
                     &Temperature (eV)            &$68.1^{+0.5}_{-0.5}$&$63.1^{+1}_{-0.8}$& $84.6^{+0.4}_{-0.4}$ &$79.9^{+1.2}_{-1.2}$\\
                     &Luminosity($10^{39}$ erg/s)&1.37&0.43&1.57&0.06\\\hline

   CIE     & Norm ($10^{58} \rm cm^{-3}$)&$0.97^{+0.47}_{-0.37}$&$1.14^{+0.71}_{-0.40}$&$0.42^{+0.13}_{-0.05}$&$0.55^{+0.04}_{-0.04}$\\
              &Temperature (keV)         &$0.18^{+0.02}_{-0.01}$&$0.17^{+0.02}_{-0.02}$&$0.19^{+0.01}_{-0.01}$&$0.26^{+0.01}_{-0.01}$\\
              & $v_{\rm mic}$ (km\ s$^{-1}$)           &2914&2914& $2914^{+612}_{-510}$&$22410^{+1688}_{-1466}$\\
              & Ne Abundance              &2.45&2.45& $2.45^{+1.36}_{-0.92}$&$2.4^{+0.9}_{-0.9}$\\
              & Mg Abundance              &1.2&1.2& $1.2^{+1.0}_{-0.6}$&$1.98^{+0.70}_{-0.70}$\\
              &Luminosity($10^{36}$ erg\ s$^{-1}$)&1.34&1.49&1.18&0.88\\ \hline

   HOT-1 & $N_H$ ($10^{24} \rm cm^{-2}$)       &$0.08^{+0.01}_{-0.01}$&$0.13^{+0.26}_{-0.05}$& $0.06^{+0.01}_{-0.01}$&$0.16^{+0.69}_{-0.02}$\\
              &Temperature (keV)                      &$0.95^{+0.04}_{-0.04}$&$2.48^{+0.52}_{-0.35}$& $0.94^{+0.02}_{-0.02}$&$1.19^{+0.86}_{-0.20}$\\
              & $\sigma_v$ rms velocity (km\ s$^{-1}$) &$892^{+43}_{-41}$&$634^{+131}_{-156}$& $932^{+23}_{-23}$&$133^{+31}_{-28}$\\
              & velocity shift (km\ s$^{-1}$)                   &$-3304^{+83}_{-83}$&$-3805^{+229}_{-278}$& $-3250^{+51}_{-51}$&$-2826^{+88}_{-97}$\\ \hline

   HOT-2 & $N_H$ ($10^{24} \rm cm^{-2}$)       &$0.76^{+0.02}_{-0.02}$&$0.44^{+0.05}_{-0.24}$&$1.61^{+0.01}_{-0.01}$&$0.71^{+0.01}_{-0.14}$\\
                 &Temperature (keV)                      &$0.89^{+0.05}_{-0.04}$&$0.59^{+0.02}_{-0.1}$& $1.06^{+0.02}_{-0.02}$&$0.68^{+0.01}_{-0.19}$\\
              & $\sigma_v$ rms velocity (km\ s$^{-1}$) &$<37$&$<20$& $35^{+14}_{-18}$&$<17$\\
              & velocity shift (km\ s$^{-1}$)                     &$-3309^{+84}_{-103}$&$-3073^{+161}_{-191}$& $-3334^{+38}_{-36}$&$-3660^{+75}_{-65}$\\\hline

   XABS & $N_H$ ($10^{20} \rm cm^{-2}$)       &$2.68^{+0.4}_{-0.3}$&$3.85^{+40}_{-2.56}$&$2.79^{+0.1}_{-0.1}$&$0.56^{+0.06}_{-0.06}$\\
                 &Log $\xi$ (erg\ cm\ s$^{-1}$)        &$2.04^{+0.14}_{-0.16}$&$3.64^{+1.35}_{-0.2}$& $2.18^{+0.05}_{-0.05}$&$0.41^{+0.08}_{-0.07}$\\
              & $\sigma_v$ rms velocity (km\ s$^{-1}$) &$49^{+126}_{-18}$&$<15831$& $53^{+10}_{-9}$& $195^{+78}_{-58}$\\
              & velocity shift (km\ s$^{-1}$)                     &$-3013^{+100}_{-368}$&$-1093^{+1294}_{-343}$& $-2968^{+47}_{-44}$&$-3127^{+127}_{-118}$\\\hline

  Abundances & C Abundance            &0.38&0.38& $0.38^{+0.07}_{-0.07}$&0.38\\
                       & N Abundance            &5.8&5.8& $5.8^{+0.2}_{-0.2}$&5.8\\
                       & O Abundance            &15.9&15.9& $15.9^{+0.5}_{-0.5}$&15.9\\
                       & Si Abundance            &0.06&0.06& $<0.06$& 0.06\\
                       & S Abundance             &0.02&0.02& $0.02^{+0.02}_{-0.01}$&0.02\\
                       & Ar Abundance            &0.5&0.5& $<0.5$& 0.5\\
                       & Ca Abundance           &0.02&0.02& $<0.02$& 0.02\\
                       & Fe Abundance           &8.9&8.9& $8.9^{+3.8}_{-2.1}$&8.9\\\hline

   HOT-ISM & $N_H$ ($10^{21} \rm cm^{-2}$)       &$3.81^{+0.05}_{-0.05}$&$4.15^{+0.05}_{-0.06}$& $3.31^{+0.02}_{-0.02}$&$2.4^{+0.1}_{-0.1}$\\
                   &Temperature (eV)                    &$1.09^{+0.07}_{-0.3}$&$1.20^{+0.03}_{-0.05}$& $1.01^{+0.04}_{-0.04}$&$0.8^{+0.1}_{-0.1}$\\
                   & N Abundance                          &1.12&1.12 & $1.12^{+0.13}_{-0.14}$&1.12\\
                   & O Abundance                          &1.68&1.68 & $1.68^{+0.02}_{-0.02}$&1.68\\
                   & Fe Abundance                         &0.81&0.81& $0.81^{+0.13}_{-0.12}$&0.81\\\hline

   AMOL (Dust) &$O_2$ ($10^{17} \rm cm^{-2}$)            & $0.47^{+0.61}_{-0.03}$& $1.48^{+0.23}_{-0.24}$&$<0.4$&$0.9^{+0.1}_{-0.1}$\\
                           &$H_2O (ice)$ ($10^{17} \rm cm^{-2}$) & $5.1^{+1.2}_{-0.1}$&$7.4^{+0.5}_{-0.5}$ &4.65$^{+0.3}_{-0.7}$ &$6.4^{+0.2}_{-0.2}$\\
                           &$CO$ ($10^{17} \rm cm^{-2}$)             &$ 1.34^{+0.53}_{-0.03}$&$0.29^{+0.2}_{-0.2}$ & $0.16^{+0.07}_{-0.06}$&$<0.2$\\\hline

                           &\rchisq         & 1.73 & 2.38 & 2.46 & 2.06 \\
                                        & (d.o.f.) & (1411) & (1435) & (1436) & (1463) \\\hline

\end{tabular}
\end{table*}

\section{Discussion}

In a nova explosion, convection and radiation pressure leads to the ejection of the envelope
material, forming an optically thick/thin shell which the high energy photons
produced by the nuclear burning have to travel through.
The resultant spectrum is an atmospheric spectrum originating from the
photosphere with a blackbody-like continuum and superimposed  absorption lines
as detected by the X-ray grating data (aside from the hard X-ray component originating from the
ejecta/winds).
In the X-ray spectral analysis of novae data, first attempts to derive spectral parameters
were done using blackbody models indicating/yielding super-Eddington luminosities
for the stellar remnant \citep[see,][]{1993Natogel}. In order to
overcome the super-Eddington luminosity problem and to incorporate the abundance and
ionization edge
effects of C-O and O-Ne core WDs, LTE (Local Thermodynamic Equilibrium)
models were used with the \emph{ROSAT} and \emph{Beppo-Sax} data and successful
results were attained \citep[e.g.,][]{1998ApJbalman, 1999AAkahabka, 2001MNRASbalman}.
However, these data had crude spectral resolution, and
only after obtaining stellar remnant spectra with the X-ray gratings,
the very detailed structure with emission and absorption features
were recovered. After this,
the few attempts to derive spectral
parameters from the stellar remnants have been done using hydrostatic NLTE atmosphere models
\citep{2002MNRASorio, 2008ApJnelson, 2010ApJrauch, 2011ApJosborne, 2011ApJness, 2013ApJtofflemire}\ 
that would account for
the absorption edges/lines from a static atmosphere.
However, the detailed structure in the spectra was very difficult to model
and results yielded best approximations to the observed spectra with large
\rchisq\ values in the fitting process.
In addition, most absorption features showed blueshifts in the spectra
\citep[e.g.,][]{2003ApJness, 2007ApJness}\ not modeled by the NLTE static atmosphere models
where only the static photospheric absorption lines were taken into account (for the analysis
of the spectra using static atmospheres, the spectra themselves 
had to be shifted for a fixed blueshift).

In order to compensate for the inadequacies, the stellar atmosphere code PHOENIX
\citep{1999JCoAMhaus, 2004AAhaus}\ 
was adjusted to model NLTE expanding atmosphere models, a hybrid atmosphere model
that is hydrostatic at the base with an expanding envelope on the top. Expansion is attained by
an optically thick wind from the remnant. These models have been used to fit the
detailed spectral data with yet again best approximations
yielding estimated spectral parameters
\citep[e.g.,][]{2005AApetz, 2010ANvanrossum}. \citet{2012ApJvanrossum}\ presents a 
new set of expanding NLTE atmosphere models with better quality fit approximations but yet
with solar composition models. One important outcome of this
modeling was that dynamic models resulted in lower effective temperatures
for the remnant WD in comparison with the static models.
This is largely due to existing edges and lines being too strong
in the static case than the expanding envelope case where features were weak. 
This can cause the static models
to measure higher temperatures than normal in order to account for the
harder X-ray tails in the observed spectra  \citep{2010ANvanrossum, 2012ApJvanrossum}.

Our approach is to mainly model the absorption components in the
X-ray spectroscopic data. This is an independent approach regarding the choice of
the continuum emission with the atmosphere models being expanding or static. We are assuming a
simple continuum model, a blackbody emission model together with a plasma emission model (collisional equilibrium),
so that the complicated absorption models are thoroughly investigated.
We note that a blackbody model does not include absorption features that could exist
in the stellar atmosphere emission, thus we will be missing some of the atmospheric absorption features.
 We note that SPEX software does not have user defined models and thus, we have no means of publicly
adding atmosphere models to use within the software.
We model the blueshifted absorption features in the line of sight as in a
nova wind or expanding ejecta.
 We infer, for this case of V2491 Cyg, two main collisionally ionized hot absorbers and an additional photoionized component
along with absorption from interstellar neutral hydrogen column density of cold gas and dust (separately)
in the line of sight.
We compare and contrast our results with the spectral parameters from
different modeling that exists in the literature, particularly with \citet{2012AApinto}\ that has analyzed
the same data using only photoionized absorbers. 

In comparison with the modeling using three photoionized absorption components
used by  \citet{2012AApinto}, our fits yield better \rchisq\
(with similar degrees of freedom) for all the regions 1, 2, 3 and the second observation (see Tables 1 \& 2).
Our findings on the plasma continuum emission parameters using the {\tt Cie} model, and the
absorption properties using the {\tt Hot-ISM}, and  {\tt Amol} dust absorption models are
similar to the results of  \citet{2012AApinto}. We take that the difference in
the quality of the fits has to do with the assumption of the utilized intrinsic absorption models.
Thus, we suggest that the major absorber in V2491 Cyg are associated with the shocked and
collisionally ionized nova ejecta and the wind which would also be consistent with the large velocity
blueshifts and turbulent line broadening, we obtained from our fits.  An additional photoionized warm 
absorber model improves the fits indicating that there is also some contribution to the complex absorption
from photoionized gas in the expanding shell/ejecta. 

The best fit model for V2491 Cyg using optically thick wind models
to fit the X-ray, near infrared, and optical light curves assumes
1.3$\pm$0.02 M$_{\odot}$ for a chemical composition of
X=0.20, Y=0.48, X$_{CNO}$=0.2, X$_{Ne}$=0.1, Z=0.02
\citep{2010ApJhachisu}.
The fits with the blackbody emission model to the \swi\ data yield
photospheric temperatures on day 39 of 36-40 eV (4.3-4.8$\times 10^5$ K), day 53 of 73-83 eV (8.8-10.0$\times 10^5$ K) and
day 80-109 of 76-81 eV (9.1-10.0$\times 10^5$ K) \citep{2010MNRASpage}.
The \XMM\ data used in our work are obtained on day 40 and 50 for which  we derive
a blackbody temperature of 61-91 eV (7.3-10.1$\times 10^5$ K) and 69-85 eV (8.3-10.0$\times 10^5$ K), respectively. These ranges have remained the same, but narrowed down slightly to 62-85 eV for day 40 and 78-82
to day 50 when an additional {\tt xabs} model was introduced.
The value (blackbody temperature) for day 40 is larger than the value obtained from \swi\ on day 40, but is in accordance with the
\swi\ value obtained for day 53. Our effective temperature values are larger compared with 
the effective temperature of 6$\times 10^5$ K obtained from the best fit
approximations to the \XMM\  data (day 50) using expanding NLTE atmosphere models
\citep{2010ANvanrossum}. In comparison with the fits using NLTE static atmosphere models where
effective temperatures of (9.6-10.5)$\times 10^5$ K are achieved \citep{2011ApJness}, our temperatures
are lower including the best fit values. The temperature ranges obtained using the fits with photoionized absorbers
are 81-123 eV for day 40 and 94-96 eV  (9.6-10.5$\times 10^5$ K) for day 50 \citep{2012AApinto}\ which are high, in general. 
Our fitted results with the two {\tt Hot} models for a collisionally ionized (in equilibrium) absorber
yield abundances of  C/C$_{\odot}$=0.3-0.5,  O/O$_{\odot}$=28-43, and  N/N$_{\odot}$=5-7. When we introduce the third photoionized absorber 
component, the abundance of the carbon and nitrogen remain
the same, but the oxygen abundance becomes O/O$_{\odot}$=15-17. 
These values are determined from material in absorption. We find that carbon is subsolar
and N supra-solar consistent with the effects of H-burning (i.e., depletion of C and enhancement of N).
The high oxygen content hints at the existence of an underlying C-O WD.  
Munari et al. (2011) calculated elemental
abundances of V2491 Cyg using ground-based optical observations. They found  Fe/Fe$_{\odot}$=0.6, O/O$_{\odot}$=4.3, N/N$_{\odot}$=59, 
and Ne/Ne$_{\odot}$=6.5. 
\citet{2010ANvanrossum}\  have only assumed solar abundances for simplicity while working with the
NLTE expanding atmosphere models and \citet{2011ApJness}\ have found consistent fits using static 
NLTE atmosphere models with fixed abundances where oxygen abundance was in a range 
 O/O$_{\odot}$=10-30. 
This latter model has similar abundances to what we derived in this study. 

Aside from the effective temperatures of the blackbody emission, we can derive the effective radius of the
blackbody emitting region to infer the WD radius. The normalization of the blackbody model fits in Table
1 \& 2 is equal to 4$\pi$R$^2$. As a result, the effective radius limits are in a range  (32-39)$\times 10^{8}$ cm 
for the region1 data, (12-29)$\times 10^{8}$ cm for the regions 2, 3 on day 40 and (2.8-7.5)$\times 10^{8}$ cm on day 50 (within the 90\% 
confidence error ranges). The addition of a third absorber component {\tt xabs}, slightly modifies these ranges to
(27-30)$\times 10^{8}$ cm
for the region1 data, (17-20)$\times 10^{8}$ cm for the regions 2, 3 on day 40 and (2.8-4.9)$\times 10^{8}$ cm on day 50.
The progressive change (getting smaller in time) of radius by a factor of about 5-10 (within errors) in about 10 days is notable. 
Such a fast decrease in effective 
stellar remnant radius by about a factor of 2-3 within a week was detected for the moderately fast nova V1974 Cyg (N Cyg 1992) 
\citep{1998ApJbalman}. \citet{1998ApJbalman}\ also notes radii in excess of 1$\times 10^{9}$ cm hinting at a bloated WD
during the early H-burning phases of the outburst. The effective blackbody temperatures that are in a range 61-91 eV (taking the largest range from Tables 1 \& 2) on day 40 and
50 constrain the WD mass in a range 1.15-1.3 M$_{\odot}$ consistent with the fast nature of the nova. 
For this calculation we assumed that the temperature range is the maximum 
temperature achieved by the H-burning WD and utilized the equation (3) in \citet{1998ApJbalman}\ that relates
maximum effective temperature with WD mass.
We underline that the radius range for the stellar remnant, on day 50, shows the desired range for the radius of a C-O WD which is
(4.5-2.8)$\times 10^{8}$ cm for the mass range of 1.15-1.3 M$_{\odot}$ \citep{1961ApJhamada, 2000AApanei}. 
Our model fits predict super-Eddington luminosities (see Table 1 \& 2)
as found in all blackbody model fits, in general. The SPEX assumes a given distance to the source fixed
in the fitting procedure, 10.5 kpc \citep{2011AAdarnley}; a smaller distance will decrease these luminosity values further. 
A recent distance estimate toward this source is 2.1-3.5 kpc \citep{2016MNRASozdonmez}\ which yields a decrease by a factor 
of 10-25 in the luminosities, reducing our values to the Eddington limit.    

We find an interstellar neutral hydrogen column density (in the
line of sight, using the {\tt Hot-ISM} model) 
for V2491 Cyg of about (3.2-4.2)$\times 10^{21}$ cm$^{-2}$ for day 40 and 
(2.2-2.7)$\times 10^{21}$ cm$^{-2}$ for day 50 from our fits (see Tables 1 \& 2). We used the 
{\tt colden} (http://cxc.harvard.edu/toolkit/colden.jsp) and {\tt nhtot} (http://www.swift.ac.uk/analysis/nhtot/) interactive program 
to calculate the  column density of hydrogen in the direction of V2491 Cyg.
The {\tt colden} program utilizes \citet{1992ApJSstark} and \citet{1990ARAAdickey}
and  {\tt nhtot}, \citet{2013MNRASwillingale} databases. The resulting values are (3.8-5.5)$\times 10^{21}$ cm$^{-2}$, in very good
agreement with our findings. Our values are also similar to
the values measured from the \swi\ data  \citep{2010MNRASpage}. 
Note that the {\tt Amol} model fits measure the dust content in the line of sight towards
the nova in a range  (0.1-2.5)$\times 10^{17}$ cm$^{-2}$ for the oxygen molecule and  (4.64-8.3)$\times 10^{17}$ cm$^{-2}$ for the water ice.
The amount of dust absorption seems to correlate with the variation seen in the LC of the first observation where the 
low count rate region2 among region1, region2 and region3 show more dust absorption. 
The second observation that has the lowest count rate has similar dust absorption to
region2 as far as the oxygen molecule and water ice are concerned.
The two collisionally ionized (in equilibrium) hot absorber components show equivalent column density of hydrogen in a range
 (0.6-5.0)$\times 10^{23}$ cm$^{-2}$ for day 40 and (1.7-3.0)$\times 10^{23}$ cm$^{-2}$ for day 50 for the first  {\tt Hot} component 
(see Table 1). 
We find progressively less absorption in this component from region1 to region3 of the LC (within errors) for the fits presented in Table 1. 
The second {\tt Hot} component
yields  (1.6-18)$\times 10^{23}$ cm$^{-2}$ for day 40 and (4.6-5.3)$\times 10^{23}$ cm$^{-2}$ for day 50. 
The second  {\tt Hot} component has less column density at the lowest count rate region2 and the second observation (see Table 1).
Once we include the additional {\tt xabs} model in the fitting procedure as displayed in Table 2,
the ranges of column density of the first {\tt Hot} 
component diminishes to (0.05-0.4)$\times 10^{23}$ cm$^{-2}$ for day 40 and (0.14-0.9)$\times 10^{23}$ cm$^{-2}$
for day 50. The second  {\tt Hot} component shows equivalent hydrogen absorption in a range (0.2-1.6)$\times 10^{23}$ cm$^{-2}$ for day 40 and 
(0.6-0.8)$\times 10^{23}$ cm$^{-2}$ for day 50 which are also lower than the fitted parameters in Table 1. 
There does not seem to be a strong pattern of variation of the range of equivalent column density of hydrogen for the {\tt Hot} components
over the four spectra fitted. There is some progressive lessening of the column density for the first  {\tt Hot} component in Table 1 for day 40.
Table 2 indicates that this component shows a higher column density by only about a factor of 2 for the data sets of region2 and the
second observation where the count rates are the lowest. The second {\tt hot} component does not indicate a particular trend.
The additional warm absorber component indicates the complexity of the absorption processes and shows that some photoionized
absorption is also at work in the expanding shell/ejecta. Table 2 shows that the photoionized absorption is  (1.3-4.3)$\times 10^{20}$ cm$^{-2}$ 
for day 40 and (0.5-0.7)$\times 10^{20}$ cm$^{-2}$ for day 50. These values are less than the interstellar column density of the hydrogen gas in the
line of sight and they comprise about (1-0.1)\%\ absorption relative to the collisionally ionized gas in absorption. 

As a result, we suggest that the large scale variation of the LC of the first observation is due to the changes in the blackbody component
parameters and intrinsic to the stellar remnant source itself. However, the changes in the absorption of the {\tt Hot-1} component
should also be noted. As the effective temperatures increase and/or the effective radius is large,
the count rates increase and when the radius and temperature decrease and/or change, the count rates change as the flux is altered 
(see Table 1 \& 2). Note that \citet{2011PASJtakei}\ found that accretion is re-established by day 50 (second observation)
and even plausibly by day 40  \citep[see also][]{2010MNRASpage}. 
Some effects of rekindled sporadic accretion onto the surface of the H-burning WD
may cause the variations seen in the LC on day 40 (first observation). Moreover, we have recovered a second blackbody 
emission component in the spectral fitting of the second observation on day 50 with a blackbody temperature of 120-131 eV  and an emitting
effective radius of (1.3-1.8)$\times 10^{7}$ cm calculated from the normalization of the fitted model (see section[3.2]). Inclusion of this component 
in the fits (in Table 2) for the second observation on day 50 diminishes the \rchisq\ to 1.97. The effective radius is consistent with  structures on an accretion disk or is about 10\%\ of the size of the WD calculated from the fits for day 50 which may be consistent 
with accretion hot spots on intermediate polar CVs.  
The temperature and emitting radius of the blackbody components are a 
support for the magnetic nature of the system as revealed by  
\citet{2015ApJzemko}\ and that accretion could have been established by day 50 \citep{2011PASJtakei}.  

We find self-consistent moving absorber velocities from 
blueshifts in a range 3097-3812 km s$^{-1}$ ({\tt Hot-1}) and  2845-3250 km s$^{-1}$ ({\tt Hot-2}) for day 40 (see Table 1). 
Table 2 shows similar ranges of absorber velocities for this component. On day 50
the two collisionally ionized absorber fits yield velocities 3517-3586 km s$^{-1}$ ({\tt Hot-1}) and 2590-2683 km s$^{-1}$ ({\tt Hot-2})
via fitting all the absorption lines simultaneously.   
These ranges are slightly altered for the second observation as 2730-2914 km s$^{-1}$ 
({\tt Hot-1}) and 3660-3735 km s$^{-1}$ ({\tt Hot-2})
when the additional {\tt xabs} model is added. 
These velocities 
are in accordance with the wind/ejecta expansion velocities (see Introduction).
We note that absorption from slower material is evident in the second observation on day 50.
The additional photoionized absorber has 
velocities calculated from the blueshift of the lines in the global fit in a range 
2645-3113 km s$^{-1}$ for
region 1, 750-2387 km s$^{-1}$ for region 2 and 2924-3015 km s$^{-1}$ for 
region 3 (day 40). On day 50 the range is 3009-3254  km s$^{-1}$. The location is similarly in the shell/ejecta
given the derived velocities in Table 2. Assuming the definition $\xi = L /n~ r^{2}$ with Eddington luminosities
and using the equivalent hydrogen densities
in Table 2, the absorber can be located (i.e., $r$) in the expanding shell/ejecta for all four fits (see also section[3.1]). 

The blueshifts measured from simple fits to
individual lines are found to be in excess of 3000 km s$^{-1}$ \citep{2011ApJness}. 
In addition, \citet{2011MNRASriberio}\ determine two components 
in the ejected material via fitting the HST (\emph{Hubble Space Telescope}) imaging data. They find
polar blobs with speeds (3100-3600) km s$^{-1}$ and an equatorial ring/belt  with (2600-2700)  km s$^{-1}$ for the first 100 d of the outburst.
These are consistent with our blueshifted {\tt Hot-1} and {\tt Hot-2}  absorber component velocities where the contribution from the 
equatorial belt/ring becomes more evident by day 50.

In order to assess the origins of the absorbers (collisionally ionized or photoionized), 
we need to consider the
ejecta or the wind by day 40 and 50 for V2491 Cyg.
The plasma temperature obtained from the \swi\ data on day 39
is 1.0-2.5 keV (1.2-3.0$\times$10$^7$ K)  and on day 53, it is 1.8-4.0 keV (2.2-4.8$\times$10$^7$ K)
\citep[for details of the {\it Swift} results see][]{2010MNRASpage}.
These values indicate origin of X-ray emission in shocked fast moving ejecta. The two {\tt Hot} 
components of the collisionally ionized absorber models show temperatures at around
1.1-3.6 keV (day 40) and 1.3-2.0 keV (day 50) for the first {\tt Hot} component. A temperature
range of  0.5-1.0 keV (day 40)
and 0.4-0.9 keV (day 50) are derived for the second {\tt Hot} component  
(Table 1 and 2 show similar/overlapping ranges). These temperatures are consistent with 
the collisionally ionized hot shocked fast moving ejecta as detected by \swi.
Note that hard X-ray component of V2491 Cyg as detected by \swi\ is
modeled by only a single component and existence of different X-ray emission components
for the hard component is not considered due to moderate spectral resolution.
For example, there can be a component from wind-wind interactions
or circumstellar interaction with different temperatures and evolution.
Also, a component arising from the
wind driven mass-loss may exist (e.g., shocks within winds due to instabilities) with 
lower X-ray temperatures and luminosity. It is important to clarify that the 0.2 keV temperature
we derive for the collisionally ionized plasma continuum emission ({\tt Cie}) could easily be
a component within the hard X-ray component.
A 0.2 keV temperature is consistent
with shock temperatures for stellar winds \citep[e.g., $\le$1 keV ($\le$1.2$\times$10$^7$ K) in general:][]{ 
1999ApJowocki}. However, the luminosity of this continuum component is large compared
with the \swi\ hard X-ray component that has a maximum at  
3$\times 10^{34}$ ergs s$^{-1}$ \citep{2010MNRASpage}. This may be because the fitted blackbody component
yields in super-Eddington luminosities with plausibly inadequate hard X-ray tail to model the spectrum 
and as a result the modeling does not predict
the right luminosities.  

We find that the turbulent velocity broadening is in a range 740-897 km s$^{-1}$ (day 40)
and 100-132 km s$^{-1}$ (day 50) for the first {\tt Hot} component.
This value is 9-62 km s$^{-1}$ (day 40) and 52-67 km s$^{-1}$ (day 50) for the second 
{\tt Hot} absorber model component. The ranges are similar when an additional {\tt xabs} model
is added to the fits for the first {\tt Hot} absorber model and the ranges are
 $<$50 km s$^{-1}$ (day 40) and $<$17 km s$^{-1}$ (day 50) for the second {\tt Hot} absorber component.
The turbulent velocity broadening ranges for the third additional photoionized absorber component are  31-174 km s$^{-1}$ for
region 1, not well constrained for region 2 and 44-63 km s$^{-1}$ for region 3 (day 40). On day 50
the range is 137-273 km s$^{-1}$. 
The values for turbulent velocity broadening suggest that the locations of the first and second
collisionally ionized absorber are different. It is an indication that the flow (of the ejecta/wind)
is inhomogeneous with locations of different densities, turbulent characteristics and temperatures. 

 

We stress that the X-ray absorption in classical/recurrent novae spectra during the outburst stage
is complicated with several different components like photospheric, 
warm photoionized absorption
and hot collisionally ionized absorption originating from a nova wind/ejecta along with
the interstellar absorption from gas and dust in the
line of sight towards the system.
Using stellar atmosphere models with photospheric absorption, expanding atmospheres or 
photoionized absorber models alone, 
as used in the literature, have been 
inadequate for modeling of the high resolution X-ray spectra, thus modeling needs to be improved.
We plan to extend our analysis to other existing data on novae and possible super soft X-ray sources to 
study how 
complex absorption affects X-ray spectra and how  
the stellar continuum is shaped during the course of the outburst evolution. 
We also aim to utilize plausible 
different continuum models as in stellar atmosphere models. 


\acknowledgements 
The Authors thank an anonymous referee for his/her critical reading of the manuscript
and valuable remarks that has improved it.
SB and CG acknowledge support
from T\"UB\.ITAK, The Scientific and Technological Research Council
of Turkey,  through project 114F351.
 

   

\begin{thebibliography}{68}
\expandafter\ifx\csname natexlab\endcsname\relax\def\natexlab#1{#1}\fi

\bibitem[{{Arnaud}(1996)}]{1996arnaud}
{Arnaud}, K.~A. 1996, in Astronomical Society of the Pacific Conference Series,
  Vol. 101, Astronomical Data Analysis Software and Systems V, ed. G.~H.
  {Jacoby} \& J.~{Barnes}, 17

\bibitem[{{Balman} \& {Krautter}(2001)}]{2001MNRASbalman}
{Balman}, {\c S}. \& {Krautter}, J. 2001, \mnras, 326, 1441

\bibitem[{{Balman} {et~al.}(1998){Balman}, {Krautter}, \&
  {{\"O}gelman}}]{1998ApJbalman}
{Balman}, {\c S}., {Krautter}, J., \& {{\"O}gelman}, H. 1998, \apj, 499, 395

\bibitem[{{Bode} \& {Evans}(2008)}]{2008bookbode}
{Bode}, M.~F. \& {Evans}, A. 2008, {Classical Novae}

\bibitem[{{Chomiuk} {et~al.}(2014){Chomiuk}, {Nelson}, {Mukai}, {Sokoloski},
  {Rupen}, {Page}, {Osborne}, {Kuulkers}, {Mioduszewski}, {Roy}, {Weston}, \&
  {Krauss}}]{2014ApJchomiuk}
{Chomiuk}, L., {Nelson}, T., {Mukai}, K., {et~al.} 2014, \apj, 788, 130

\bibitem[{{Darnley} {et~al.}(2011){Darnley}, {Ribeiro}, {Bode}, \&
  {Munari}}]{2011AAdarnley}
{Darnley}, M.~J., {Ribeiro}, V.~A.~R.~M., {Bode}, M.~F., \& {Munari}, U. 2011,
  \aap, 530, A70

\bibitem[{{den Herder} {et~al.}(2001){den Herder}, {Brinkman}, {Kahn},
  {Branduardi-Raymont}, {Thomsen}, {Aarts}, {Audard}, {Bixler}, {den Boggende},
  {Cottam}, {Decker}, {Dubbeldam}, {Erd}, {Goulooze}, {G{\"u}del}, {Guttridge},
  {Hailey}, {Janabi}, {Kaastra}, {de Korte}, {van Leeuwen}, {Mauche},
  {McCalden}, {Mewe}, {Naber}, {Paerels}, {Peterson}, {Rasmussen}, {Rees},
  {Sakelliou}, {Sako}, {Spodek}, {Stern}, {Tamura}, {Tandy}, {de Vries},
  {Welch}, \& {Zehnder}}]{2001AAdenherder}
{den Herder}, J.~W., {Brinkman}, A.~C., {Kahn}, S.~M., {et~al.} 2001, \aap,
  365, L7

\bibitem[{{Dickey} \& {Lockman}(1990)}]{1990ARAAdickey}
{Dickey}, J.~M. \& {Lockman}, F.~J. 1990, \araa, 28, 215

\bibitem[{{Ferland}(2003)}]{2003ARAAferland}
{Ferland}, G.~J. 2003, \araa, 41, 517

\bibitem[{{Hachisu} \& {Kato}(2001)}]{2001ApJhachisu}
{Hachisu}, I. \& {Kato}, M. 2001, \apj, 558, 323

\bibitem[{{Hachisu} \& {Kato}(2010)}]{2010ApJhachisu}
{Hachisu}, I. \& {Kato}, M. 2010, \apj, 709, 680

\bibitem[{{Hamada} \& {Salpeter}(1961)}]{1961ApJhamada}
{Hamada}, T. \& {Salpeter}, E.~E. 1961, \apj, 134, 683

\bibitem[{{Hauschildt} \& {Baron}(1999)}]{1999JCoAMhaus}
{Hauschildt}, P.~H. \& {Baron}, E. 1999, Journal of Computational and Applied
  Mathematics, 109, 41

\bibitem[{{Hauschildt} \& {Baron}(2004)}]{2004AAhaus}
{Hauschildt}, P.~H. \& {Baron}, E. 2004, \aap, 417, 317

\bibitem[{{Helton} {et~al.}(2008){Helton}, {Woodward}, {Vanlandingham}, \&
  {Schwarz}}]{2008CBEThelton}
{Helton}, L.~A., {Woodward}, C.~E., {Vanlandingham}, K., \& {Schwarz}, G.~J.
  2008, Central Bureau Electronic Telegrams, 1379

\bibitem[{{Henze} {et~al.}(2014){Henze}, {Ness}, {Darnley}, {Bode}, {Williams},
  {Shafter}, {Kato}, \& {Hachisu}}]{2014AAhenze}
{Henze}, M., {Ness}, J.-U., {Darnley}, M.~J., {et~al.} 2014, \aap, 563, L8

\bibitem[{{Hernanz} \& {Sala}(2002)}]{2002Scihernanz}
{Hernanz}, M. \& {Sala}, G. 2002, Science, 298, 393

\bibitem[{{Hernanz} \& {Sala}(2007)}]{2007ApJhernanz}
{Hernanz}, M. \& {Sala}, G. 2007, \apj, 664, 467

\bibitem[{{Ibarra} {et~al.}(2009){Ibarra}, {Kuulkers}, {Osborne}, {Page},
  {Ness}, {Saxton}, {Baumgartner}, {Beckmann}, {Bode}, {Hernanz}, {Mukai},
  {Orio}, {Sala}, {Starrfield}, \& {Wynn}}]{2009AAibarra}
{Ibarra}, A., {Kuulkers}, E., {Osborne}, J.~P., {et~al.} 2009, \aap, 497, L5

\bibitem[{{Jansen} {et~al.}(2001){Jansen}, {Lumb}, {Altieri}, {Clavel}, {Ehle},
  {Erd}, {Gabriel}, {Guainazzi}, {Gondoin}, {Much}, {Munoz}, {Santos},
  {Schartel}, {Texier}, \& {Vacanti}}]{2001AAjansen}
{Jansen}, F., {Lumb}, D., {Altieri}, B., {et~al.} 2001, \aap, 365, L1

\bibitem[{{Kaastra} {et~al.}(1996){Kaastra}, {Mewe}, \&
  {Nieuwenhuijzen}}]{1996kaastra}
{Kaastra}, J.~S., {Mewe}, R., \& {Nieuwenhuijzen}, H. 1996, in UV and X-ray
  Spectroscopy of Astrophysical and Laboratory Plasmas, ed. K.~{Yamashita} \&
  T.~{Watanabe}, 411--414

\bibitem[{{Kahabka} {et~al.}(1999){Kahabka}, {Hartmann}, {Parmar}, \&
  {Negueruela}}]{1999AAkahabka}
{Kahabka}, P., {Hartmann}, H.~W., {Parmar}, A.~N., \& {Negueruela}, I. 1999,
  \aap, 347, L43

\bibitem[{{Kallman} \& {Bautista}(2001)}]{2001ApJSkallman}
{Kallman}, T. \& {Bautista}, M. 2001, \apjs, 133, 221

\bibitem[{{Krautter} {et~al.}(1996){Krautter}, {Oegelman}, {Starrfield},
  {Wichmann}, \& {Pfeffermann}}]{1996ApJkrautter}
{Krautter}, J., {Oegelman}, H., {Starrfield}, S., {Wichmann}, R., \&
  {Pfeffermann}, E. 1996, \apj, 456, 788

\bibitem[{{Livio}(1994)}]{1994inbilivio}
{Livio}, M. 1994, in Saas-Fee Advanced Course 22: Interacting Binaries, ed.
  S.~N. {Shore}, M.~{Livio}, E.~P.~J. {van den Heuvel}, H.~{Nussbaumer}, \&
  A.~{Orr}, 135--262

\bibitem[{{Lynch} {et~al.}(2008){Lynch}, {Russell}, {Rudy}, {Woodward}, \&
  {Schwarz}}]{2008IAUClynch}
{Lynch}, D.~K., {Russell}, R.~W., {Rudy}, R.~J., {Woodward}, C.~E., \&
  {Schwarz}, G.~J. 2008, \iaucirc, 8935

\bibitem[(1994)]{l2} Lodders, K. \& Palme, H., 2009, in 72nd Annual Meeting of
the Meteoritical Society, Meteoritics and Planetary Science Supplement ser., 72, 5154

\bibitem[{{MacDonald} \& {Vennes}(1991)}]{1991ApJmacdonald}
{MacDonald}, J. \& {Vennes}, S. 1991, \apjl, 373, L51

\bibitem[{{Mukai} \& {Ishida}(2001)}]{2001ApJmukai}
{Mukai}, K. \& {Ishida}, M. 2001, \apj, 551, 1024

\bibitem[{{Munari} {et~al.}(2011){Munari}, {Siviero}, {Dallaporta}, {Cherini},
  {Valisa}, \& {Tomasella}}]{2011NEWAmunari}
{Munari}, U., {Siviero}, A., {Dallaporta}, S., {et~al.} 2011, \na, 16, 209

\bibitem[{{Nakano} {et~al.}(2008){Nakano}, {Beize}, {Jin}, {Gao}, {Yamaoka},
  {Haseda}, {Guido}, {Sostero}, {Klingenberg}, \& {Kadota}}]{2008IAUCnakano}
{Nakano}, S., {Beize}, J., {Jin}, Z.-W., {et~al.} 2008, \iaucirc, 8934

\bibitem[{{Nelson} {et~al.}(2008){Nelson}, {Orio}, {Cassinelli}, {Still},
  {Leibowitz}, \& {Mucciarelli}}]{2008ApJnelson}
{Nelson}, T., {Orio}, M., {Cassinelli}, J.~P., {et~al.} 2008, \apj, 673, 1067

\bibitem[{{Ness} {et~al.}(2009){Ness}, {Drake}, {Starrfield}, {Bode},
  {O'Brien}, {Evans}, {Eyres}, {Helton}, {Osborne}, {Page}, {Schneider}, \&
  {Woodward}}]{2009AJness}
{Ness}, J.-U., {Drake}, J.~J., {Starrfield}, S., {et~al.} 2009, \aj, 137, 3414

\bibitem[{{Ness} {et~al.}(2011){Ness}, {Osborne}, {Dobrotka}, {Page}, {Drake},
  {Pinto}, {Detmers}, {Schwarz}, {Bode}, {Beardmore}, {Starrfield}, {Hernanz},
  {Sala}, {Krautter}, \& {Woodward}}]{2011ApJness}
{Ness}, J.-U., {Osborne}, J.~P., {Dobrotka}, A., {et~al.} 2011, \apj, 733, 70

\bibitem[{{Ness} {et~al.}(2007){Ness}, {Starrfield}, {Beardmore}, {Bode},
  {Drake}, {Evans}, {Gehrz}, {Goad}, {Gonzalez-Riestra}, {Hauschildt},
  {Krautter}, {O'Brien}, {Osborne}, {Page}, {Sch{\"o}nrich}, \&
  {Woodward}}]{2007ApJness}
{Ness}, J.-U., {Starrfield}, S., {Beardmore}, A.~P., {et~al.} 2007, \apj, 665,
  1334

\bibitem[{{Ness} {et~al.}(2003){Ness}, {Starrfield}, {Burwitz}, {Wichmann},
  {Hauschildt}, {Drake}, {Wagner}, {Bond}, {Krautter}, {Orio}, {Hernanz},
  {Gehrz}, {Woodward}, {Butt}, {Mukai}, {Balman}, \& {Truran}}]{2003ApJness}
{Ness}, J.-U., {Starrfield}, S., {Burwitz}, V., {et~al.} 2003, \apjl, 594, L127

\bibitem[{{O'Brien} {et~al.}(1994){O'Brien}, {Lloyd}, \&
  {Bode}}]{1994MNRASobrien}
{O'Brien}, T.~J., {Lloyd}, H.~M., \& {Bode}, M.~F. 1994, \mnras, 271, 155

\bibitem[{{Oegelman} {et~al.}(1993){Oegelman}, {Orio}, {Krautter}, \&
  {Starrfield}}]{1993Natogel}
{Oegelman}, H., {Orio}, M., {Krautter}, J., \& {Starrfield}, S. 1993, \nat,
  361, 331

\bibitem[{{Orio} {et~al.}(2003){Orio}, {Hartmann}, {Still}, \&
  {Greiner}}]{2003ApJorio}
{Orio}, M., {Hartmann}, W., {Still}, M., \& {Greiner}, J. 2003, \apj, 594, 435

\bibitem[{{Orio} {et~al.}(2001){Orio}, {Parmar}, {Benjamin}, {Amati},
  {Frontera}, {Greiner}, {{\"O}gelman}, {Mineo}, {Starrfield}, \&
  {Trussoni}}]{2001MNRASorio}
{Orio}, M., {Parmar}, A., {Benjamin}, R., {et~al.} 2001, \mnras, 326, L13

\bibitem[{{Orio} {et~al.}(2002){Orio}, {Parmar}, {Greiner}, {{\"O}gelman},
  {Starrfield}, \& {Trussoni}}]{2002MNRASorio}
{Orio}, M., {Parmar}, A.~N., {Greiner}, J., {et~al.} 2002, \mnras, 333, L11

\bibitem[{{Orio} {et~al.}(2015){Orio}, {Rana}, {Page}, {Sokoloski}, \&
  {Harrison}}]{2015MNRASorio}
{Orio}, M., {Rana}, V., {Page}, K.~L., {Sokoloski}, J., \& {Harrison}, F. 2015,
  \mnras, 448, L35

\bibitem[{{Osborne} {et~al.}(2011){Osborne}, {Page}, {Beardmore}, {Bode},
  {Goad}, {O'Brien}, {Starrfield}, {Rauch}, {Ness}, {Krautter}, {Schwarz},
  {Burrows}, {Gehrels}, {Drake}, {Evans}, \& {Eyres}}]{2011ApJosborne}
{Osborne}, J.~P., {Page}, K.~L., {Beardmore}, A.~P., {et~al.} 2011, \apj, 727,
  124

\bibitem[{{Owocki} \& {Cohen}(1999)}]{1999ApJowocki}
{Owocki}, S.~P. \& {Cohen}, D.~H. 1999, \apj, 520, 833

\bibitem[{{{\"O}zd{\"o}nmez} {et~al.}(2016){{\"O}zd{\"o}nmez}, {G{\"u}ver},
  {Cabrera-Lavers}, \& {Ak}}]{2016MNRASozdonmez}
{{\"O}zd{\"o}nmez}, A., {G{\"u}ver}, T., {Cabrera-Lavers}, A., \& {Ak}, T.
  2016, \mnras, 461, 1177

\bibitem[{{Page} {et~al.}(2010){Page}, {Osborne}, {Evans}, {Wynn}, {Beardmore},
  {Starling}, {Bode}, {Ibarra}, {Kuulkers}, {Ness}, \&
  {Schwarz}}]{2010MNRASpage}
{Page}, K.~L., {Osborne}, J.~P., {Evans}, P.~A., {et~al.} 2010, \mnras, 401,
  121
\bibitem[{{Page} {et~al.}(2015){Page}, {Osborne}, {Kuin}, {Henze}, {Walter},
  {Beardmore}, {Bode}, {Darnley}, {Delgado}, {Drake}, {Hernanz}, {Mukai},
  {Nelson}, {Ness}, {Schwarz}, {Shore}, {Starrfield}, \&
  {Woodward}}]{2015MNRASpage}
{Page}, K.~L., {Osborne}, J.~P., {Kuin}, N.~P.~M., {et~al.} 2015, \mnras, 454,
  3108

\bibitem[{{Panei} {et~al.}(2000){Panei}, {Althaus}, \&
  {Benvenuto}}]{2000AApanei}
{Panei}, J.~A., {Althaus}, L.~G., \& {Benvenuto}, O.~G. 2000, \aap, 353, 970

\bibitem[{{Petz} {et~al.}(2005){Petz}, {Hauschildt}, {Ness}, \&
  {Starrfield}}]{2005AApetz}
{Petz}, A., {Hauschildt}, P.~H., {Ness}, J.-U., \& {Starrfield}, S. 2005, \aap,
  431, 321

\bibitem[{{Pinto} {et~al.}(2010){Pinto}, {Kaastra}, {Costantini}, \&
  {Verbunt}}]{2010AApinto}
{Pinto}, C., {Kaastra}, J.~S., {Costantini}, E., \& {Verbunt}, F. 2010, \aap,
  521, A79

\bibitem[{{Pinto} {et~al.}(2012){Pinto}, {Ness}, {Verbunt}, {Kaastra},
  {Costantini}, \& {Detmers}}]{2012AApinto}
{Pinto}, C., {Ness}, J.-U., {Verbunt}, F., {et~al.} 2012, \aap, 543, A134

\bibitem[{{Rauch} {et~al.}(2010){Rauch}, {Orio}, {Gonzales-Riestra}, {Nelson},
  {Still}, {Werner}, \& {Wilms}}]{2010ApJrauch}
{Rauch}, T., {Orio}, M., {Gonzales-Riestra}, R., {et~al.} 2010, \apj, 717, 363

\bibitem[{{Ribeiro} {et~al.}(2011){Ribeiro}, {Darnley}, {Bode}, {Munari},
  {Harman}, {Steele}, \& {Meaburn}}]{2011MNRASriberio}
{Ribeiro}, V.~A.~R.~M., {Darnley}, M.~J., {Bode}, M.~F., {et~al.} 2011, \mnras,
  412, 1701

\bibitem[{{Shara}(1989)}]{1989PASPshara}
{Shara}, M.~M. 1989, \pasp, 101, 5

\bibitem[{{Sokoloski} {et~al.}(2006){Sokoloski}, {Luna}, {Mukai}, \&
  {Kenyon}}]{2006Natursokoloski}
{Sokoloski}, J.~L., {Luna}, G.~J.~M., {Mukai}, K., \& {Kenyon}, S.~J. 2006,
  \nat, 442, 276

\bibitem[{{Stark} {et~al.}(1992){Stark}, {Gammie}, {Wilson}, {Bally}, {Linke},
  {Heiles}, \& {Hurwitz}}]{1992ApJSstark}
{Stark}, A.~A., {Gammie}, C.~F., {Wilson}, R.~W., {et~al.} 1992, \apjs, 79, 77

\bibitem[{{Starrfield} {et~al.}(2012){Starrfield}, {Iliadis}, {Timmes}, {Hix},
  {Arnett}, {Meakin}, \& {Sparks}}]{2012BASIstarrfield}
{Starrfield}, S., {Iliadis}, C., {Timmes}, F.~X., {et~al.} 2012, Bulletin of
  the Astronomical Society of India, 40, 419

\bibitem[{{Str{\"u}der} {et~al.}(2001){Str{\"u}der}, {Briel}, {Dennerl},
  {Hartmann}, {Kendziorra}, {Meidinger}, {Pfeffermann}, {Reppin}, {Aschenbach},
  {Bornemann}, {Br{\"a}uninger}, {Burkert}, {Elender}, {Freyberg}, {Haberl},
  {Hartner}, {Heuschmann}, {Hippmann}, {Kastelic}, {Kemmer}, {Kettenring},
  {Kink}, {Krause}, {M{\"u}ller}, {Oppitz}, {Pietsch}, {Popp}, {Predehl},
  {Read}, {Stephan}, {St{\"o}tter}, {Tr{\"u}mper}, {Holl}, {Kemmer}, {Soltau},
  {St{\"o}tter}, {Weber}, {Weichert}, {von Zanthier}, {Carathanassis}, {Lutz},
  {Richter}, {Solc}, {B{\"o}ttcher}, {Kuster}, {Staubert}, {Abbey}, {Holland},
  {Turner}, {Balasini}, {Bignami}, {La Palombara}, {Villa}, {Buttler},
  {Gianini}, {Lain{\'e}}, {Lumb}, \& {Dhez}}]{2001AAstruder}
{Str{\"u}der}, L., {Briel}, U., {Dennerl}, K., {et~al.} 2001, \aap, 365, L18

\bibitem[{{Takei} {et~al.}(2011){Takei}, {Ness}, {Tsujimoto}, {Kitamoto},
  {Drake}, {Osborne}, {Takahashi}, \& {Kinugasa}}]{2011PASJtakei}
{Takei}, D., {Ness}, J.-U., {Tsujimoto}, M., {et~al.} 2011, \pasj, 63, S729

\bibitem[{{Tofflemire} {et~al.}(2013){Tofflemire}, {Orio}, {Page}, {Osborne},
  {Ciroi}, {Cracco}, {Di Mille}, \& {Maxwell}}]{2013ApJtofflemire}
{Tofflemire}, B.~M., {Orio}, M., {Page}, K.~L., {et~al.} 2013, \apj, 779, 22

\bibitem[{{Tomov} {et~al.}(2008){Tomov}, {Mikolajewski}, {Ragan},
  {Swierczynski}, \& {Wychudzki}}]{2008ATeltomov}
{Tomov}, T., {Mikolajewski}, M., {Ragan}, E., {Swierczynski}, E., \&
  {Wychudzki}, P. 2008, The Astronomer's Telegram, 1475

\bibitem[{{Turner} {et~al.}(2001){Turner}, {Abbey}, {Arnaud}, {Balasini},
  {Barbera}, {Belsole}, {Bennie}, {Bernard}, {Bignami}, {Boer}, {Briel},
  {Butler}, {Cara}, {Chabaud}, {Cole}, {Collura}, {Conte}, {Cros}, {Denby},
  {Dhez}, {Di Coco}, {Dowson}, {Ferrando}, {Ghizzardi}, {Gianotti}, {Goodall},
  {Gretton}, {Griffiths}, {Hainaut}, {Hochedez}, {Holland}, {Jourdain},
  {Kendziorra}, {Lagostina}, {Laine}, {La Palombara}, {Lortholary}, {Lumb},
  {Marty}, {Molendi}, {Pigot}, {Poindron}, {Pounds}, {Reeves}, {Reppin},
  {Rothenflug}, {Salvetat}, {Sauvageot}, {Schmitt}, {Sembay}, {Short},
  {Spragg}, {Stephen}, {Str{\"u}der}, {Tiengo}, {Trifoglio}, {Tr{\"u}mper},
  {Vercellone}, {Vigroux}, {Villa}, {Ward}, {Whitehead}, \&
  {Zonca}}]{2001AAturner}
{Turner}, M.~J.~L., {Abbey}, A., {Arnaud}, M., {et~al.} 2001, \aap, 365, L27

\bibitem[{{van Rossum}(2012)}]{2012ApJvanrossum}
{van Rossum}, D.~R. 2012, \apj, 756, 43

\bibitem[{{van Rossum} \& {Ness}(2010)}]{2010ANvanrossum}
{van Rossum}, D.~R. \& {Ness}, J.-U. 2010, Astronomische Nachrichten, 331, 175

\bibitem[{{Verner} {et~al.}(1996){Verner}, {Verner}, \&
  {Ferland}}]{1996ADNDTverner}
{Verner}, D.~A., {Verner}, E.~M., \& {Ferland}, G.~J. 1996, Atomic Data and
  Nuclear Data Tables, 64, 1

\bibitem[{{Verner} \& {Yakovlev}(1995)}]{1995AASverner}
{Verner}, D.~A. \& {Yakovlev}, D.~G. 1995, \aaps, 109

\bibitem[{{Webbink} {et~al.}(1987){Webbink}, {Livio}, {Truran}, \&
  {Orio}}]{1987ApJwebbink}
{Webbink}, R.~F., {Livio}, M., {Truran}, J.~W., \& {Orio}, M. 1987, \apj, 314,
  653

\bibitem[{{Willingale} {et~al.}(2013){Willingale}, {Starling}, {Beardmore},
  {Tanvir}, \& {O'Brien}}]{2013MNRASwillingale}
{Willingale}, R., {Starling}, R.~L.~C., {Beardmore}, A.~P., {Tanvir}, N.~R., \&
  {O'Brien}, P.~T. 2013, \mnras, 431, 394

\bibitem[{{Zemko} {et~al.}(2015){Zemko}, {Mukai}, \& {Orio}}]{2015ApJzemko}
{Zemko}, P., {Mukai}, K., \& {Orio}, M. 2015, \apj, 807, 61

\end{thebibliography}
\end{document}